\documentclass[a4paper,11pt]{article}
\usepackage{jheppub} 
\usepackage{lineno}
\usepackage[utf8]{inputenc}

\usepackage{jheppub}
\usepackage{graphicx,verbatim}
\usepackage{blindtext}
\usepackage[T1]{fontenc}
\usepackage{epsf}
\usepackage{mathtools}
\usepackage{lmodern}
\usepackage[mathscr]{euscript}

\usepackage{ytableau}
\usepackage{youngtab}

\usepackage{amsmath}
\usepackage{amsfonts}
\usepackage{amssymb}
\usepackage{mathrsfs}
\usepackage{slashed}
\usepackage{xcolor}

\def\kq{\mathfrak{q}}
\def\fq{\mathfrak{q}}

\def\fg{\mathfrak{g}}

\def\rx{\mathrm{x}}
\def\ry{\mathrm{y}}
\def\rz{\mathrm{z}}

\def\ri{\mathrm{i}}

\def\bi{\mathbf{i}}
\def\bj{\mathbf{j}}
\def\bz{\mathbf{z}}

\def\ba{\mathbf{a}}

\def\bb{\mathbf{b}}

\def\bS{\mathbf{S}}
\def\bN{\mathbf{N}}
\def\bK{\mathbf{K}}

\def\bX{\mathbf{X}}

\def\BC{\mathbb{C}}
\def\BR{\mathbb{R}}
\def\BE{\mathbb{E}}
\def\BZ{\mathbb{Z}}

\def\BB{\mathbb{B}}

\def\CalN{\mathcal{N}}

\def\CalR{\mathcal{R}}
\def\CalZ{\mathcal{Z}}

\def\CalW{\mathcal{W}}

\def\CalO{\mathcal{O}}
\def\CalM{\mathcal{M}}

\def\CalT{\mathcal{T}}

\def\CalL{\mathcal{L}}
\def\CalX{\mathcal{X}}
\def\Tr{{\rm Tr}}

\def\ve{{\varepsilon}}
\def\bbZ{\mathbb{Z}}

\def\tx{{\tilde x}}

\def\ti{\mathtt{i}}

\def\tn{\mathtt{n}}
\def\tx{\mathtt{x}}
\def\tp{\mathtt{p}}
\def\tq{\mathtt{q}}

\def\fz{\mathfrak{z}}

\def\EK{\EuScript{K}}

 \def\p{\partial}

 \def\d{\delta}

 \def\k{\kappa}

 \def\s{\sigma}

 \def\o{\omega }

\def\beq{\begin{equation}}
\def\eeq{\end{equation}}

\title{Generalized Calogero-Moser system and supergroup gauge origami}

\author[{\spadesuit}]{Taro Kimura} 
\author[{\clubsuit}]{, Norton Lee}
\affiliation[{\spadesuit}]{Institut de Mathématiques de Bourgogne, Université de Bourgogne, CNRS, France\footnote{%
Unité Mixte de Recherche (UMR 5584) commune au CNRS et à l'Université de Bourgogne}}
\affiliation[{\clubsuit}]{Center for Geometry and Physics, Institute for Basic Science (IBS), Pohang 37673, Republic of Korea}

\emailAdd{taro.kimura@u-bourgogne.fr}
\emailAdd{norton.lee@ibs.re.kr}

\preprint{CGP24006}

\abstract{We study the integrability and the Bethe/Gauge correspondence of the Generalized Calogero-Moser system proposed by Berntson, Langmann and Lenells \cite{Berntson:2023prh} which we call the elliptic quadruple Calogero-Moser system (eqCM). 
We write down the Dunkl operators which give commuting Hamiltonians of the quantum integrable system.  
We identify the gauge theory in correspondence is a supergroup version of the gauge origami, from which we construct the transfer matrix of the eqCM system. 
}

\begin{document}
\maketitle
\flushbottom

\section{Introduction}\label{sec:intro}

The intrinsic connection between the 4d $\CalN=2$ supersymmetric gauge theories and algebraic integrable system has been an intense interest and fruitful research ground since the groundbreaking work of Seiberg and Witten through the identification between the Seiberg-Witten curve of the supersymmetric gauge theory and the spectral curve of the algebraic integrable system \cite{Seiberg:1996nz,Seiberg:1994aj,Seiberg:1994rs}.
Nekrasov and Shatashvili promoted the relation to between the quantum integrable system and 4d $\CalN=2$ symmetric gauge theory subject to $\Omega$-deformation. The $\Omega$-deformation allows localization computation of the supersymmetric partition function and other protected observables. This was afterwards named as \emph{Bethe/Gauge correspondence}. 
In general, there can be multiple deformation parameters $\{\ve_a\}$ in the localization colputation based on the dimensionality of the supersymmetric gauge theory considered. 
It is proposed that the $\Omega$-deformation parameters $\{\ve_a\}$ can be regarded in general as Planck constants when quantizing the classical integrable system. 
In the Nekrasov-Shatashvili limit where one of the $\Omega$-deformation parameter is turned off, an effective $\CalN=(2,2)$ supersymmetry is restored on the corresponding two dimensional subspace. 
Simultaneously, the Bethe ansatz equation (BAE) of the quantum integrable system is recovered from the saddle point equation of the supersymmetric gauge theory partition function. 
In this relationship the effective twisted superpotential of the $\CalN=(2,2)$ gauge theory corresponds to the Yang-Yang functional of the quantum integrable system. In this regard, the stationary state of the quantum integrable systems are the vacuua of the effective $\CalN=(2,2)$ theory.  
The quantum Hamiltonians are identified with the generator of the chiral rings \cite{Nekrasov:2009uh,NRS2011,Nikita-Shatashvili,HYC:2011, Dorey:2011pa}. Well-known examples include $\CalN=2$ SYM/Toda lattice \cite{Martinec:1995by}, $\CalN=2$ SQCD/XXX spin chain-Gaudin model \cite{Lee:2020hfu,HYC:2011,Jeong:2018qpc}, and $\CalN=2^*$/Calogero-Moser system \cite{DW1,DP1}.  

Recent studies show that on top of the $\Omega$-deformation and taking the appropriate limit \cite{Nekrasov:2012xe}, a co-dimensional two full type surface defect needs to be introduced through appropriate orbifolding procedures \cite{Kanno:2011fw,Nakajima:2011yq} to fully understand the relation between the conserving quantum Hamiltonian and (twisted) chiral ring operator \cite{Nekrasov:2017gzb,Lee:2020hfu}. 
Introduction of such full type surface defect generates the chainsaw quiver gauge theory on the two dimensional surface. 
For a $G$-gauge theory, the quiver gauge theory is equipped with a set of gauge couplings $(\kq_i)_{i=1,\dots {\rm rk}G}$, which would be identified as the canonical coordinates in the quantum integrable system. 
This allows us to construct the quantum conserved Hamiltonians from gauge theory~\cite{Chen:2019vvt}, and the further identification between instanton partition function and quantum wave function \cite{Nekrasov:2017gzb,Jeong:2023qdr,jeong2021intersecting,Jeong:2024mxr,Jeong:2024hwf}. 

\subsection*{Supergroup and superalgebra}

The supergroup and underlying superalgebra is well known to be an extension of Lie group and Lie algebra by including Grassmannian generators. 
In the study of supersymmetric quantum field theory, the supergroup often encodes both the bosonic and fermionic degrees of freedom of the global spacetime symmetry. 
However, quantum field theory with supergroup as its gauge group is inevitably non-unitary due to the violation of spin-statistics theorem, leaving these class of theories to be often over-looked and unexplored in the literature. 
The recent advance in computational techniques brings back the interests of studying these exotic theories. 
Most notably, the explicit constructions of supergroup gauge theory via D-branes and so-called negative/ghost D-branes were given in \cite{Okuda:2006fb, Dijkgraaf:2016lym,Nekrasov:2018xsb} (See \cite{Vafa:2001qf} also for earlier work), moreover the  partition functions for supergroup gauge instantons have been computed in \cite{Kimura:2019msw,Kimura:2023iup} using equivariant localization. 

Algebraic integrable system such as Calogero-Moser system  can be defined with any given Lie algebra (see \cite{Ruijsenaars:1999PF} for a good introduction). 
The supermatrix provide natural generalization of the Lax-formalism of integrable system to superalgebraic integrable system \cite{Evans:1996bu}, including super Toda lattice \cite{van1994super}, superspin chain \cite{Beisert:2003yb,Nekrasov:2018gne} , and superanalog of Calogero-Moser system \cite{veselov1996new,sergeev2004deformed, sergeev2001superanalogs}. 

It is natural to investigate if the well-established Bethe/Gauge correspondence extends to their corresponding super version. 
At the classical level, 
we can naturally generalize the Lax operators/matrices of ordinary integrable systems to construct the so-called super-Lax operators which take values on the supermatrices. 
The super analog of the Calogero-Moser (CM) system is the double Calogero--Moser (dCM) system with both trigonometric version (trigonometric dCM; tdCM)~\cite{HiJack, sergeev2001superanalogs} and elliptic version (elliptic dCM; edCM)~\cite{sergeev2004deformed,Sergeev:2017jqy}.
Integrability, which is again defined by the existence of many commuting conserved Hamiltonians, of the dCM system was proved in \cite{HiJack} for the trigonometric, and in \cite{Chen:2019vvt} for the elliptic cases.
There are two gauge theories in correspondence to the edCM system. One is 4d $\CalN=2^*$ supergroup gauge theory \cite{Chen:2020rxu}. The other one is the \emph{folded instanton} \cite{Chen:2019vvt, Nekrasov:2017gzb} based on the gauge origami \cite{Nikita:III,Nikita:IV}. 

In this paper we study the integrability and Bethe/Gauge correspondence of a generalization of the edCM system, which we will call the elliptic quadruple Calogero-Moser system (eqCM), proposed in \cite{Berntson:2023prh}. It is a 1+1 dimensional system of arbitrary number of four kinds of particles. Our goal is to prove its integrability  by explicit construct its Dunkl operator, and establish its Bethe/Gauge correspondence with supergroup folded instanton.

\subsection*{Outline}

In section \ref{sec:qCM} we review the eqCM system and prove its integrability by explicitly constructing its Dunkl operator. 
In section \ref{sec:supergroup}, we will review the D-brane realization of the supergroup gauge theory and study the gauge origami with supergroup. 
In section \ref{sec:Bethe} we will explicitly construct the supergroup folded instanton partition function, and derive the BAE from from its saddle point equation. 
We recover the quantum Hamiltonian of the eqCM system by introducing full type surface defect. The last part is achieved through the regularity of $qq$-character observable \cite{Nikita:I}. 
Finally we discuss various future directions in section \ref{sec:discussion}.
We relegate our various definitions of functions and some of the computational details in a series of Appendices.

\acknowledgments

We would like to thank Edwin Langman for fruitful discussion at the workshop ``\href{https://sites.google.com/view/msj-si-2023/home}{The 16th MSJ-SI: Elliptic Integrable Systems, Representation Theory and Hypergeometric Functions}'' held at Tokyo, July-August 2023, from which this work arises.
We are grateful to the workshop organizers for providing such a stimulating atmosphere.
The work of TK was supported by CNRS through MITI interdisciplinary programs, EIPHI Graduate School (No.~ANR-17-EURE-0002) and Bourgogne-Franche-Comté region.
The work of NL is supported by IBS project IBS-R003-D1.

\section{Quadruple Calogero-Moser system} \label{sec:qCM}

The elliptic Calogero-Moser (eCM) system is defined for any finite-dimenaional Lie algebra $\fg$ by the Hamiltonian \cite{olshanetsky1983quantum}:
\begin{align}
    {\rm H}_\fg = - \frac{1}{2} \Delta_\fg + \sum_{\alpha \in \CalR^+(\fg)} g_\alpha (g_\alpha-1) (\alpha,\alpha) \wp(\alpha\cdot \rx) 
\end{align}
where $\CalR^+(\fg)$ is the set of all positive roots of Lie algebra $\fg$, $\Delta_\fg$ is the Laplace operator defined on the Lie algebra $\fg$.
The Weierstrass $\wp$-function $\wp(z;2\ell,2\ri\d)=\wp(z)$ has half-period $(\ell,\ri\d)$. See also~\eqref{eq:cmplx_coupling}. $g_\alpha$ is the potential coupling that depends on the length of the fundamental root $\alpha$. 
The eCM system defined based on the Lie algebra $A_{N-1}$, which is a one-dimensional quantum mechanics system of $N$ particles with the Hamiltonian:
\begin{align}\label{def:eCM}
    {\rm H}_{N;g}(\rx) = - \sum_{j=1}^N \frac{\p^2}{\p\rx_j^2} + g(g-1) \sum_{1\leq j < k \leq N} \wp(\rx_j-\rx_k), 
\end{align}
is closely related to four dimensional $\CalN=2^*$ $SU(N)$ gauge theory.

The algebraic integrable system can be defined on the Lie superalgebra. Let $I = I_{\bar{0}} \amalg I_{\bar{1}}$ be the union of "even" indices $I_{\bar{0}} = \{1,\dots,N\}$ and "odd" indices $I_{\bar{1}} = \{\bar{1},\dots,\bar{M}\}$. 
Let $V = \BC^{N|M}$ and $e_1,\dots,e_N,{e}_{\bar{1}},\dots,{e}_{\bar{M}}$ be its basis such that each vector's parity be aligned with its index. 
Let $e^*_1,\dots,e^*_N,e^*_{\bar{1}},\dots,e^*_{\bar{M}}$ be the basis of the dual space $V^*$. Then the root system of the Lie superalgebra $\fg = \mathfrak{gl}(N|M)$ can be described as follows: $\CalR = \amalg_{\sigma,\sigma' = \pm} \CalR_{\sigma \sigma'}
$ where
\begin{align}
\begin{split}
    & \CalR_{++} = \{e^*_i-e^*_j | \ i,j \in I_{\bar{0}} \}, \quad  \CalR_{+-} = \{e^*_i-e^*_j | \ i\in I_{\bar{0}}, \ j \in I_{\bar{1}} \}, \\
    & \CalR_{-+} = \{e^*_i-e^*_j | \ i \in I_{\bar{1}}, \ j \in I_{\bar{0}},  \}, \quad  \CalR_{--} = \{e^*_i-e^*_j | \ i,j \in I_{\bar{1}} \}.
\end{split}
\end{align}
On the dual space $V^*$, define inner product that depends on a parameter $g$ by
\begin{align}\label{def:inner-product}
    (v^*,w^*)_g = \sum_{i=1}^N v^*(e_i) w^*(e_i) - g \sum_{j=1}^M v^*(e_{\bar{j}}) w^*(e_{\bar{j}}) .
\end{align}

The elliptic Calogero-Moser system associated with the root system of the superalgebra $\mathfrak{g} = \mathfrak{gl}(N|M)$, called elliptic double Calogero-Moser (edCM) system, is defined by: 
\begin{align}\label{def:edCM}
    {\rm H}_{N,M;g}(\rx,\ry) = {\rm H}_{N;g}(\rx) - g {\rm H}_{M;1/g}(\ry) + \sum_{j=1}^N \sum_{k=1}^M (1-g) \wp (\rx_j-{\ry}_k) .
\end{align}
The integrability of the edCM system was proven in \cite{Chen:2019vvt}. It is related to two gauge theories: the four dimensional $\CalN=2^*$ $SU(N|M)$ supergroup gauge theory \cite{Chen:2020rxu} and folded instanton \cite{Chen:2019vvt,Nekrasov:2017gzb}.  

Bernston, Langmann and Lenells proposed a generalization of the edCM system in \cite{Berntson:2023prh}. 
The generalized eCM model, which we call elliptic quadruple Calogero-Moser system (eqCM) here, is a one-dimensional quantum mechanics system of $L=N_++N_-+M_++M_-$ particles governed by the Hamiltonian:
\begin{align}\label{def:ddCM}
\begin{split}
    & {\rm H}_{N_+,M_+,N_-,M_-;g} (\rx^+,\ry^+,\rx^-,{\ry}^-;g) \\
    & = {\rm H}_{N_+,M_+;g} (\rx^+,{\ry}^+) + {\rm H}_{N_-,M_-;g}(\rx^-,{\ry}^-) \\
    & \quad + \sum_{j=1}^{N_+} \sum_{k=1}^{N_-} g (g-1) \wp (\rx_j^+-\rx_k^-+\ri\d) + \sum_{j=1}^{M_+} \sum_{k=1}^{M_-} \left( 1 - \frac{1}{g} \right) \wp({\ry}^+_j - {\rx}_k^- + \ri \d) \\
    & \quad + \sum_{j=1}^{N_+} \sum_{k=1}^{M_-} (1-g) \wp (\rx^+_j-{\ry}^-_k + \ri \d) + \sum_{j=1}^{N_-} \sum_{k=1}^{M_+} (1-g) \wp ({\ry}^+_j - \rx^-_k + \ri \d) .
\end{split}
\end{align}
The eqCM \eqref{def:ddCM} describes a 1+1 dimensional system with four kinds particles interact through two-body interaction based on their physical distance $|x|$. 
The particles can be labeled $(m_j,c_j) = (1,+)$, $(1,-)$, $(1/g,+)$, and $(1/g,-)$. 
The label $m_j$ characterizes the mass of the particles. The two types of particles of different mass corresponds to electrons and holes respectively in the condense matter system \cite{berntson2020nonchiral}. 
The label $c_j = \pm$ can be realized as chirality of the particles, corresponding to the left and right moving particles.    
The particles of the same chirality interact through the potential proportional to $\wp(x)$, while the particles of opposite chirality interact through potential proportional to $\wp(x+\ri \d)$. The former is singular when $x \to 0$ and is repulsive, while the later is non-singular in $x \to 0$ and is attractive
\begin{equation}
    \lim_{x\to 0} \wp(x) = \frac{(\pi/\d)^2}{\sinh^2\frac{\pi x}{\d} }, \quad \lim_{x\to 0} \wp(x+\ri\d) = -\frac{(\pi/\d)^2}{\cosh^2\frac{\pi x}{\d} }. 
\end{equation}
%


\subsection{Integrability}
Before we dive into the gauge theory construction of the eqCM system, we would like to investigate its integrability discussed in \cite{Berntson:2023prh}. 
It turns out that the integrability of the eqCM system solely depends on the integrability of the edCM. The proof is straightforward: By shifting $\ry^\pm \to \ry^\pm + \ri\d$ the eqCM becomes an edCM. 
The integrability of the edCM has been proved in \cite{Chen:2019vvt}. Hence, the eqCM system is integrable. 

We construct the Dunkl operator for the eqCM system by considering a family of functions 
\begin{align}
    \sigma_t(x) = \frac{\theta_{11}(x) \theta'_{11}(0) }{\theta_{11}(x) \theta_{11}(-t) }, \quad t \in \BC / (\BZ \oplus  \tau\BZ)
\end{align}
where $\tau$ is a modular parameter and $\theta_{11}$ is the theta function (Recall that we have identify complex gauge coupling with the elliptic modulus at the end of previous chapter).
The function $\sigma_t(x)$ has the following properties
\begin{subequations}\label{sigma properties}
	\begin{align}
    &\sigma_t(x+2\pi i)=\sigma_t(x), \\
    &\sigma_t(x)=-\sigma_{-t}(-x)\label{sigma minus}, \\
    &\sigma_t(x)=-\sigma_x(t) ,\\
    &\sigma_t(x)\sigma_{-t}(x)=\wp(x)-\wp(t), \\
    &\lim_{t\to0}\frac{d}{dx}\sigma_t(x)=-\wp(x)-2\zeta\left(\frac{1}{2}\right).
    \end{align}
\end{subequations}
For the later convenience, we will collectively denote the coordinates by $\rz_j$, $j=1,\dots,L=N_1+N_2+M_1+M_2$. 
\begin{align}
    \rz_j = \begin{cases}
        \rx_j^+ & j=1,\dots,N_+, \\
        \rx_{j-N_+}^- - \ri\d & j=N_1+1,\dots,N_++N_-, \\
        \ry_{j-N_+-N_-}^+ & j = N_++N_-+1,\dots,N_++N_-+M_+, \\
        \ry_{j-N_+-N_--M_+}^- - \ri\d & j= N_++N_- + M_++1, \dots, N_++N_-+M_++M_-
    \end{cases}
\end{align}
We define the parity function 
\begin{align}
    \tp(j) = \begin{cases}
        0 & j=1,\dots,N_++N_- \\
        1 & j=N_++N_-+1,\dots,L
    \end{cases} 
\end{align}
such that the partricles' chiralities are given by $c_j = (-1)^{\tp(j)}$. 
Let $t_j \in \BC/(\BZ \oplus \tau \BZ)$ ($j=1,\dots,L$) be $L$ complex numbers. The elliptic Dunkl operator is defined by
\begin{align}\label{def:Dunkl}
    D_j = g^{\tp(j)} \frac{\p}{\p\rz_j} + \sum_{\substack{k=1 \\ k\neq j}}^L g^{1-\tp(k)} \sigma_{t_j-t_k}(\rz_j - \rz_k) S_{jk}
\end{align}
Here $S_{jk}=S_{kj}$ are permutation operators acting on $\{\rz_j\}$ by $\rz_j S_{jk} = S_{jk} \rz_{k}$. 
Here we show the Dunkl operator defined in \eqref{def:Dunkl} are pairwise commuting: 
\begin{align}
\begin{split}
    \left[ D_j, D_k \right]
    & = \left[ g^{\tp(j)} \frac{\p}{\p\rz_j}, \sum_{\substack{l=1 \\ l\neq k}}^L g^{1-\tp(l)} \sigma_{t_k-t_l}(\rz_k-\rz_l) S_{kl} \right] + \left[ \sum_{\substack{l=1 \\ l\neq j}}^L g^{1-\tp(l)} \sigma_{t_j-t_l}(\rz_j-\rz_l) S_{jl}, g^{\tp(k)} \frac{\p}{\p\rz_k} \right] \\
    & = \left[ g^{\tp(j)} \frac{\p}{\p\rz_j}, g^{1-\tp(j)} \sigma_{t_k-t_j}(\rz_k-\rz_j) S_{kj} \right] + \left[ g^{1-\tp(k)} \sigma_{t_j-t_k}(\rz_j-\rz_k) S_{jk}, g^{\tp(k)} \frac{\p}{\p\rz_k} \right] \\
    & = \frac{\p}{\p\rz_j} \sigma_{t_k-t_j}(\rz_k-\rz_j) S_{kj} - \sigma_{t_k-t_j}(\rz_k-\rz_j) \frac{\p}{\p \rz_k}S_{kj}  \\
    & \qquad + \sigma_{t_j-t_k}(\rz_j-\rz_k) \frac{\p}{\p\rz_j} S_{jk} - \frac{\p}{\p\rz_k} \sigma_{t_j-t_k}(\rz_j-\rz_k) S_{jk} \\
    & = \left[ \frac{\p}{\p\rz_j}, \sigma_{t_k-t_j}(\rz_k-\rz_j) \right] S_{jk} + \left[ \frac{\p}{\p\rz_k}, \sigma_{t_k-t_j}(\rz_k-\rz_j) \right] S_{jk} = 0. 
\end{split}
\end{align}
The conserved charges are given by
\begin{align}\label{def:CC-dunkl}
    \CalL^{(r)} = \sum_{j=1}^L g^{-\tp(j)} (D_j)^r.  
\end{align}
In the classical integrable system, the conserving Hamiltonians are obtained by the (super)trace of the $L\times L$ Lax matrix. Eq.~\eqref{def:CC-dunkl} is the quantum uplift and generalization of the classical conserved charge. The "Lax matrix" is a diagonal $L \times L$ matrix with diagonal components $D_j$. The trace is deformed in the same manner by inner product \eqref{def:inner-product}. 

Since individual Dunkl operators $D_j$ are pairwise commuting, it is obvious that $\CalL^{(r)}$ are also pairwise commuting
\begin{align}
    \left[ \CalL^{(r)}, \CalL^{(s)} \right] = 0, \quad \forall \; r,s=1,\dots,L. 
\end{align}
In particular, to recover the eqCM Hamiltonian, we consider
\begin{align}
\begin{split}
    \CalL^{(2)} = & \sum_{j=1}^L g^{-\tp(j)} (D_j)^2 = \sum_{j=1}^{N_1+N_2} (D_j)^2 + \sum_{j=N_1+N_2+1}^L \frac{1}{g} (D_j)^2 \\
    = & \sum_{j=1}^{N_1+N_2} \frac{\p^2}{\p\rz_j^2} + g \sum_{\substack{k=1 \\ k\neq j}}^{N_1+N_2} \frac{\p}{\p\rz_j} \sigma_{t_j-t_k}(\rz_j-\rz_k) + g^2 \sum_{\substack{k=1 \\ k \neq j}}^{N_1+N_2} \sigma_{t_j-t_k}(\rz_j-\rz_k) \sigma_{t_k-t_j}(\rz_k-\rz_j) \\
    & \qquad \qquad + \sum_{k=N_1+N_2+1}^L \frac{\p}{\p \rz_j} \sigma_{t_j-t_k}(\rz_j-\rz_k) + \sum_{k=N_1+N_2+1}^L \sigma_{t_j-t_k}(\rz_j-\rz_k) \sigma_{t_k-t_j}(\rz_k-\rz_j) \\
    & + \sum_{j=N_1+N_2+1}^{L} \frac{g^2}{g} \frac{\p^2}{\p\rz_j^2} + \frac{g^2}{g} \sum_{{k=1}}^{N_1+N_2} \frac{\p}{\p\rz_j} \sigma_{t_j-t_k}(\rz_j-\rz_k) + \frac{g^2}{g} \sum_{k=1}^{N_1+N_2} \sigma_{t_j-t_k}(\rz_j-\rz_k) \sigma_{t_k-t_j}(\rz_k-\rz_j) \\
    & \qquad \qquad + \frac{g}{g} \sum_{\substack{ k=N_1+N_2+1 \\ k\neq j }}^L \frac{\p}{\p \rz_j} \sigma_{t_j-t_k}(\rz_j-\rz_k) + \frac{1}{g} \sum_{\substack{ k=N_1+N_2+1 \\ k\neq j }}^L \sigma_{t_j-t_k}(\rz_j-\rz_k) \sigma_{t_k-t_j}(\rz_k-\rz_j) \\
\end{split}
\end{align}
In the limit that all $t_j=t$ identical, we use the fourth and fifth property of $\sigma_t(x)$ in \eqref{sigma properties} to obtain the eqCM Hamiltonian
\begin{align}
    \left.\CalL^{(2)}\right|_{t_j=t} = {\rm H}_{N_+,M_+,N_-,M_-}(\rx^+,\ry^+,\rx^-,\ry^-;-g). 
\end{align}
We hence prove the integrability of the eqCM system.

\section{Supergroup gauge origami}\label{sec:supergroup}

The D-brane construction of supergroup gauge theory was first studied in~\cite{Vafa:2001qf,Okuda:2006fb}, and later studied in various other papers such as \cite{Kapustin:2009cd,Witten:2011zz,Dijkgraaf:2016lym,Mikhaylov:2014aoa,Mikhaylov:2015nsa,Okazaki:2017sbc,Aghaei:2018cbn,Kimura:2019msw,Kimura:2020lmc,Kimura:2021ngu,Kimura:2023ndz}.
The concept of \emph{negative brane (ghost brane)} is introduced to study supergroup gauge theory, which we will give a brief review here, see \cite{Vafa:2001qf,Kapustin:2009cd,Kimura:2020jxl,Kimura:2023iup} for detail. 

\subsection{Supergroup from D-branes}
The D-brane acts as the boundary on the string world sheet. In the presence of multiple D-branes, a Chan-Paton factor, which takes a value in the vector space $\BC^N$ in the case of $N$ D-branes, is assigned to each of the string to identify which D-brane it ends on. These Chan-Paton factors, though non-dynamical on the world sheet, give rise to the $N$ labels that generate the $U(N)$ gauge group in the target space. 
The Dirac quantization condition restricts $N$ to be integer. It does not, however, rule out the possibility of negative $N$. 
A negative Chan-Paton factor can be realized by introducing additional minus sign to certain subset of boundary states, i.e. the Chan-Paton factor now takes a value in a graded vector space $\mathbb{C}^{N_+|N_-}$ with the mixed signatures:
\begin{equation}\label{metric}
     \begin{pmatrix}
     +\mathbf{1}_{N_+} & 0 \\
    0 & -\mathbf{1}_{N_-}
    \end{pmatrix}.
\end{equation}
The additional vector space with negative signature corresponds to the additional $N_-$ \textit{negative branes} which carry an additional negative sign with respect to the boundary states, hence the Chan-Paton factors corresponding to the usual \textit{positive branes}. This distinguishes them from the more commonly encountered anti-branes which only carry an negative sign for the RR sector, and preserve the opposite space-time supersymmetries. A system consisting of both positive and negative branes can preserve the same space time supersymmetries. Moreover due to the negative sign in their NS-NS sector, a negative brane carries negative tension and anti-gravitates, this feature again differs from the anti-brane which gravitates. 

Let us consider the scenario of $N_+$ positive and $N_-$ negative D3 branes place in parallel. A string that connects between one positive brane and one negative brane have the opposite statistics comparing to a string having both ends on positive brane. This is caused by the additional minus sign on the negative brane. 
An open string with both ends on the negative brane will have usual statistics since the minus signs from both ends cancel each other. 
This enhances $U(N_+) \times U(N_-)$ quiver gauge group to a $U(N_+|N_-)$ supergroup,
\begin{align}
    U = \begin{pmatrix}
        A & B \\ C & D
    \end{pmatrix} \in U(N_+|N_-), \quad U^{-1} = U^\dagger. 
\end{align}
Here $A$ and $D$ are $N_+ \times N_+$ and $N_-\times N_-$ even graded matrix with complex number entries, while $B$ and $C$ are $N_+\times N_-$ and $N_- \times N_+$ odd graded matrix whose entries are Grassmannian. 
An odd graded matrix means its entries obey anti-commutation relation instead of commutation relation,
\begin{align}
    B_{ij}C_{kl} = (-1) C_{kl}B_{ij}. 
\end{align}

\subsection{Instanton moduli space of supergroup gauge origami}

Here we recall the supergroup instanton moduli space studied in \cite{Kimura:2019msw} and extend it to the gauge origami. 
We apply the supergroup D-brane construction to the gauge origami, which is inspired by type IIB string theory \cite{Nikita:III,Chen:2019vvt}. 
Let us start with four complex planes $\BC^4$ with coordinates $\bz_a$, $a=1,2,3,4$, and consider picking out two of them such that the six possible copies $\BC^2_A \subset\BC^4$ are labeled by two digit indices,
\begin{align}
    A \in \{(12),(13),(14),(23),(24),(34)\} = \underline{6}.
\end{align}
We also define 
\begin{align}
    \underline{4} = \{1,2,3,4\} .
\end{align}
Define the complement of $A$ by $\bar{A} = \underline{4}\backslash A$, for instance if $A=(12)$, $\bar{A}=(34)$. 
The complementary two complex planes transverse to $\BC^2_{A}$ is thus denoted by $\BC^2_{\bar{A}}$, and $\BC^4 = \BC^2_A \oplus \BC^2_{\bar{A}} $. 
One can imagine the four complex planes labeled by four complex coordinates $\bz_{1,2,3,4}$ are the four faces/edges of a tetrahedron and the six subspaces $\BC^2_A$ are the edges of the tetrahedron. 
This is inspired by the D1-D5-$\overline{\rm D5}$ system in \cite{NaveenNikita}. Noted that the $\overline{\rm D5}$ denotes the anti-brane instead of the negative brane. 
We consider the D(-1)-D3-$\overline{\rm D3}$ system which can be obtained through T-duality (Table~\ref{tab:GO}).

\begin{table}[h!]
    \centering
    \begin{tabular}{|c||c|c|c|c|c|c|c|c|c|c|c|}
    \hline
     IIB branes & 1 & 2 & 3 & 4 & 5 & 6 & 7 & 8 & 9 & 10 \\
     \hline
     \hline
     ${\rm D(-1)}_\pm$ & & & & & & & & & & \\
     \hline
     ${\rm D3}_{(12),\pm}$ & x & x & x & x & & & & & & \\
     \hline
     $\overline{\rm D3}_{(13),\pm}$ & x & x & & & x & x & & & & \\
     \hline
     ${\rm D3}_{(14),\pm}$ & x & x & & & & & x & x & & \\
     \hline
     ${\rm D3}_{(23),\pm}$ & & & x & x & x & x & & & & \\
     \hline
     $\overline{\rm D3}_{(24),\pm}$ & & & x & x & & & x & x & & \\
     \hline
     ${\rm D3}_{(34),\pm}$ & & & & & x & x & x & x & & \\
     \hline
    \end{tabular}
    \caption{D-brane setup for gauge origami.}
    \label{tab:GO}
\end{table}

We label the D3${}_\pm$ and $\overline{\rm D3}_\pm$ branes by ${\rm D3}_{A,\pm}$ and $\overline{\rm D3}_{A,\pm}$ indicating the four dimensional world volume is in $\BC^2_A$. The $\pm$ labels the sign of the Chan-Paton factor of the D-branes. Notice that the two out of six $\overline{\rm D3}$ branes is necessary for the interesting brane configuration to preserve $\frac{1}{16}$ of supersymmetry or 2 supercharges. 
We also introduced $(\k^+|\k^-)$ D(-1)${}_\pm$ branes, which will play the role of ``spiked instantons'' in this configuration.
Similar to ADHM construction for supergroup \cite{Kimura:2019msw}, each gauge group is associated to a graded vector space 
\[
    \bN_A=\BC^{n_A^+|n_A^-} = \BC^{n_A^+} \oplus \BC^{n_A^-} = \bN_{A,+} \oplus \bN_{A,-}
\]
and one additional vector space associated to $(\k^+|\k^-)$ D(-1) branes.
\[
    \bK=\mathbb{C}^{\k^+|\k^-} = \BC^{\k^+} \oplus \BC^{\k^-} = \bK_+ \oplus \bK_-.
\]

The analogous maps acting on $\{\bN_A\}$ and $\bK$ are:
\begin{subequations}
\begin{align}
& I_A: \bN_A\to \bK; \\
& J_A: \bK\to \bN_A; \\
& B_a: \bK\to \bK;\quad a \in \underline{4} = \{1,2,3,4\},
\end{align}
\end{subequations}
and we can understand these from the world volume theory of $(\k^+|\k^-)$ D(-1) branes.
Here $\{I_A\}$ and $\{J_A\}$ are the bi-fundamental fields arising from the open string stretching among the D3${}_{A,\pm}$/$\overline{\text{D3}}_{A,\pm}$ and D(-1) branes, while $B_a$ are the complex $U(\k^+|\k^-)$ adjoint fields whose diagonal entries label the positions of $(\k^+|\k^-)$ D(-1) branes in the transverse $\BC^4$.
The analogous real moment map $\mu_{\mathbb{R}}$ and complex moment map $\mu_{\mathbb{C}}$ equations to ADHM construction in $\mathbb{C}^4$ can thus be identified respectively with the so-called D-term, E-term and J-term conditions \cite{NaveenNikita}. Starting with the D-term, which gives the real moment map $\mu_{\BR}: \bK \to \bK$,
\begin{equation}\label{mu=0}
    \left\{ \mu_{\mathbb{R}} = \sum_{a\in\underline{4}}[B_a,B_a^\dagger]+
    \sum_{A\in\underline{6}}I_AI_A^\dagger-J_A^\dagger J_A=\zeta\cdot \text{diag} (1_{\bK_+},-1_{\bK_-}) \right\}\Big/U(\k^+|\k^-). 
\end{equation}
Here in such an intersecting D-brane configuration, we also turn on constant background NS-NS B-field, it generates the FI parameter in the D(-1) brane world volume theory.
Next we would like to discuss the E- and J-term conditions together. To do so, for each $\bN_A$, let us define the following map $\mu_A : \bK \to \bK$,
\begin{equation}
\mu_A=[B_a,B_b]+I_AJ_A;\quad a, b \in A
\end{equation} 
and we define $s_A$ as:
\begin{equation}\label{}
s_A=\mu_A+\varepsilon_{A\bar{A}}\mu^\dagger_{\bar{A}},
\end{equation}
where $\varepsilon_{A\bar{A}}$ is a four indices totally antisymmetric tensor ranging over $A$ and its complement $\bar{A}$. 
The analogue of the complex moment maps are now given by:
\begin{equation}\label{sa=0}
\{s_A=0\}/U(\k^+|\k^-).
\end{equation}
Notice that, while $\mu_A = 0$ can encode six complex equations, $s_A$ consists both $\mu_A$ and $\mu^\dagger_{\bar{A}}$ which are mapped into each other under hermitian conjugation. 
There are therefore only six real equations encoded in \eqref{sa=0}.
The reason of using $s_A$ instead of $\mu_A$ is because $s_A$ gives the correct number of degree of freedom as we will show in a moment. 
In addition, there are equations which do not exist in the usual ADHM construction:
\begin{equation}\label{BIBJ=0}
\{\sigma_{\bar{a}A}=B_{\bar{a}}I_A+\varepsilon_{\bar{a}\bar{b}} B^\dagger_{\bar{b}}J^\dagger_A=0\}/U(\k^+|\k^-):\bN_A\to \bK,
\end{equation}
where $\bar{a} \in \bar{A}$ denotes the single index contained in the double index $\bar{A}= \underline{4} \backslash A $.
For every $A$, there are two such equations. 
These equations \eqref{BIBJ=0} appear when one considers the D(-1) and intersecting D3 instanton configurations~\cite{Nikita:II}. 
The moduli space of the spiked instanton is defined by
\begin{align}\label{def:moduli}
    \mathcal{M}_{\k,n} =\{(\vec{B},\vec{I},\vec{J})\mid\eqref{mu=0},\eqref{sa=0},\eqref{BIBJ=0}\}
\end{align}
where $\k = (\k^+ | \k^-)$ and $n = \{ n_A^\sigma \}_{A \in \underline{6},\sigma = \pm}$.
By the structure of the supermatrix \eqref{eq:real-moment-supermatrix}, there are three types of solution (up to $U(\k^+|\k^-)$ transformation) to Eq.~\eqref{mu=0} based on the property of the FI-parameter $\zeta$:
\begin{itemize}
    \item If $\zeta$ is a real number, the moduli space is bosonic, i.e. $B_a$, $I_A$ and $J_A$ are block diagonal. 
    \item If $\zeta$ is a Grassmannian even number, the moduli space is fermionic, i.e. $B_a$, $I_A$ and $J_A$ are block off-diagonal. 
    \item If $\zeta$ is a combination of Grassmannian even number and real number, the moduli space is mixed.
\end{itemize}
In this note we only consider the bosonic moduli space to serve our purpose.
We would like to claim that equations in \eqref{sa=0}, \eqref{mu=0}, and \eqref{BIBJ=0} are sufficient to fix the solution uniquely by showing the number of degrees of freedom and the number of conditions are equal.
Let us start with counting the degrees of freedom. 
For bosonic moduli space: 
\begin{enumerate}
	\item $I_A$: $\displaystyle 2\sum_A \left( \k^+n_A^+ + \k^-n_A^-\right)$ real d.o.f
	\item $J_A$: $\displaystyle 2\sum_A \left( \k^+n_A^+ + \k^-n_A^-\right)$ real d.o.f.
	\item $B_a$: $4\times 2 \times((\k^+)^2+(\k^-)^2)$ real d.o.f.
\end{enumerate}
which together precisely equals to the number of the conditions:
\begin{enumerate}
	\item Eq.~\eqref{sa=0}: $6\times((\k^+)^2+(\k^-)^2)$ real conditions
	\item Eq.~\eqref{mu=0}: $(\k^+)^2+(\k^-)^2$ real conditions.
	\item Eq.~\eqref{BIBJ=0}: $\displaystyle \sum_A 2\times2\times(\k^+n_A^++\k^- n_A^-)$ real conditions
	\item $U(\k^+|\k^-)$ symmetry: $(\k^+)^2+(\k^-)^2$ real conditions.
\end{enumerate}
Hence we may show that the dimensions of moduli space $\CalM_{\k^+|\k^-}$ \eqref{def:moduli} is exactly 
\begin{align}
\begin{split}
    & 2 \sum\limits_A \k^+n_A^+ + \k^-n_A^- + 2 \sum\limits_A \k^+n_A^+ + \k^-n_A^- + 8 ((\k^+)^2+(\k^-)^2) \\
    & - 6((\k^+)^2+(\k^-)^2) - ((\k^+)^2+(\k^-)^2) - \sum_A2\times2\times(\k^+n_A^++\k^- n_A^-) - ((\k^+)^2+(\k^-)^2) \\
    & = 0 \nonumber
\end{split}
\end{align}

%
Essentially $\CalM_{\k,n}$ only consists of only discrete points.%
\footnote{This situation is analogous to the moduli space of the coherent sheaves on Calabi--Yau three-folds, whose virtual dimension becomes zero due to the Serre duality.} 
Comparing to the ADHM moduli space of $\k$-instanton configurations of $U(n)$ gauge theory denoted by $\CalM_{\k,n}$ with $\dim_\mathbb{C} \CalM_{\k,n} = 2\kappa n$, additional eq.~\eqref{BIBJ=0} reduces instanton moduli space to be zero dimensional. 
\paragraph{}
However, there also exist open strings stretching between D3-D3 branes which gives additional maps/fields from D-brane construction. These terms are not related to instanton and thus not being considered when constructing instanton moduli space. For instance, when one considers the D(-1)-D3-brane realization of ADHM construction, the open strings with both ends attached to D3 branes are not taken into account. Here we also consider the open strings stretching between $\text{D3}_A$-$\text{D3}_{\bar{A}}$ branes and $\overline{\text{D3}}_A$-$\overline{\text{D3}}_{\bar{A}}$, giving rise to the following conditions:
\begin{equation}\label{JIIJ=0}
\Upsilon_A=J_{\bar{A}}I_A-I^\dagger_{\bar{A}}J^\dagger_A=0:\bN_A\to \bN_{\bar{A}},
\end{equation}
these act as the transversality conditions \cite{Nikita:II}.
The matrices $\{s_A\}, \{\sigma_{\bar{a}A}\}$ and $\{\Upsilon_A\}$ in \eqref{sa=0}, \eqref{BIBJ=0}, and \eqref{JIIJ=0} need to be subjected to the following matrix consistency identity \cite{Nikita:II}:
\begin{align}\label{Consistency1}
\begin{split}
    & \sum_{A\in\underline{6}}\text{str}(s_As_A^\dagger)+\sum_{A\in\underline{6},\bar{a}\in\underline{4}}\text{str}(\sigma_{\bar{a}A}\sigma_{\bar{a}A}^\dagger)+\sum_{A\in\underline{6}}{\rm str}(\Upsilon_A\Upsilon_A^\dagger) \\
	& =2\sum_{A\in\underline{6}}\left(\|\mu_A\|^2+\|J_{\bar{A}}I_A \|^2\right)+\sum_{A\in\underline{6},\bar{a}\in\bar{A}}(\|B_{\bar{a}}I_A\|^2+\|J_AB_{\bar{a}}\|^2),
\end{split}
\end{align}
where $\|\mu_A\|^2=\text{str}\left(\mu_A\mu_A^\dagger\right)$. 
Define projection $\pi_\pm: \bK \to \bK_\pm$, $\pi_\pm^2 = \pi_\pm^\dagger = \pi_\pm$, $\pi_+\bK_-=0 = \pi_-\bK_+$. We can than consider the projection of \eqref{Consistency1} onto
\begin{align}
\begin{split}
    & \sum_{A\in\underline{6}}\text{str}(\pi_+ s_A\pi_+\pi_+s_A^\dagger \pi_+)+\sum_{A\in\underline{6},\bar{a}\in\underline{4}}\text{str}(\pi_+\sigma_{\bar{a}A}\pi_+\pi_+\sigma_{\bar{a}A}^\dagger\pi_+)+\sum_{A\in\underline{6}}{\rm str}(\pi_+\Upsilon_A\pi_+\pi_+\Upsilon_A^\dagger\pi_+) \\
    = & \sum_{A\in\underline{6}}\Tr_{\bK_+}(\pi_+s_A\pi_+\pi_+s_A^\dagger\pi_+)+\sum_{A\in\underline{6},\bar{a}\in\underline{4}}\Tr_{\bK_+}(\pi_+\sigma_{\pi_+\bar{a}A}\pi_+\pi_+\sigma_{\bar{a}A}^\dagger\pi_+)+\sum_{A\in\underline{6}}\Tr_{\bK_+}(\pi_+\Upsilon_A\pi_+\pi_+\Upsilon_A^\dagger\pi_+) \\
	= & 2 \sum_{A\in\underline{6}}\left(\Tr_{\bK_+}(\pi_+\mu_A\pi_+\pi_+\mu_A^\dagger\pi_+)+\Tr_{\bK_+}(\pi_+J_{\bar{A}}I_A \pi_+\pi_+ I_A^\dagger J_{\bar{A}}^\dagger\pi_+)  \right) \\
    & +\sum_{A\in\underline{6},\bar{a}\in\bar{A}}(\Tr_{\bK_+}(\pi_+B_{\bar{a}}I_A\pi_+\pi_+I_A^\dagger B_{\bar{a}}^\dagger\pi_+)+\Tr_{\bK_+}(\pi_+J_AB_{\bar{a}}\pi_+\pi_+B_{\bar{a}}^\dagger J^\dagger_A\pi_+) ),
\end{split}
\end{align}
The right RHS is non-negative. By setting each term in the LHS vanishing  using  \eqref{sa=0}, \eqref{BIBJ=0}, and \eqref{JIIJ=0}, we can deduce the following constraints:
\begin{subequations}
\begin{align}
	\{s_A=0\}/U(\k^+|\k^-)\implies & \pi_+\mu_A\pi_+=0, \\
	\{\sigma_{\bar{a}A}=0\}/U(\k^+|\k^-)\implies & \pi_+B_{\bar{a}}I_A\pi_+=0;\quad \pi_+J_AB_{\bar{a}}\pi_+=0, \\
	\{\Upsilon_A=0\}/U(\k^+|\k^-) \implies & \pi_+J_{\bar{A}}I_A\pi_+=0,
\end{align}
\end{subequations}
A similar argument holds for the $\pi_-$. Combining the two constraints gives
\begin{subequations}
\begin{align}
	\pi_\pm\mu_A\pi_\pm=0 \implies & \mu_A = 0,\label{Improv1} \\
	\pi_\pm B_{\bar{a}}I_A\pi_\pm=0; \ \pi_\pm J_AB_{\bar{a}}\pi_\pm=0 \implies & I_A\pi_\pm=0 = J_AB_{\bar{a}} ,\label{Improv2} \\
	\pi_\pm J_{\bar{A}}I_A\pi_\pm=0 \implies & J_{\bar{A}}I_A = 0\label{Improv3}
\end{align}
\end{subequations}
which are equivalent to the E- and J-term constraints considered in \cite{NaveenNikita}.
\paragraph{}
It is known that the combination of imposing $\zeta>0$ and dividing by $U(\k^+|\k^-)$ in \eqref{mu=0} is equivalent to replacing D-term equation \eqref{mu=0} by the (co-)stability condition \cite{Nikita:II,Kimura:2019msw}, which states that for any subspace $\bK'_+\subset\bK_+$, such that $I_A(\bN_{A,+})$ for all $A\in\underline{6}$ and $B_a\bK'_+\subset\bK_+$ for all $a=1,2,3,4$, coincides with $\bK_+$, i.e. $\bK'_+=\bK_+$, and for any subspace $\bK'_-\subset\bK_-$, such that $J_A^\dagger(\bN_{A,-})$ for all $A\in\underline{6}$ and $B_a^\dagger\bK'_-\subset\bK_-$ for all $a=1,2,3,4$, coincides with $\bK_-$, i.e. $\bK'_-=\bK_-$.
In other words, 
\begin{align} \label{stability}
\begin{split}
    \bK_+=\sum_A\mathbb{C}[B_1,B_2,B_3,B_4] I_A(\bN_{A,+})/ GL(\bK_+) \\
    \bK_-=\sum_A\mathbb{C}[B_1^\dagger,B_2^\dagger,B_3^\dagger,B_4^\dagger] J_A^\dagger (\bN_{A,-})/ GL(\bK_-) \\
\end{split}
\end{align}
The equations \eqref{Improv2} and \eqref{Improv3} further shows that $\bK_\pm$ can be decomposed into
\begin{align}\label{Decompose K}
\begin{split}
    \bK_+=\bigoplus_A\bK_{A,+};\quad \bK_{A,+}=\mathbb{C}[B_a,B_b]I_A(\bN_{A,+}), \\
    \bK_-=\bigoplus_A\bK_{A,-};\quad \bK_{A,-}=\mathbb{C}[B_a^\dagger,B_b^\dagger]J_A^\dagger(\bN_{A,-}),
\end{split}
\end{align}
The equation \eqref{Decompose K} is essentially the (co-)stability condition for the familiar ADHM moduli space. Combining \eqref{Improv1} and \eqref{Decompose K}, we have shown that gauge origami is actually six independent copies of ADHM construction of instantons. 
Finally, the moduli space is now isomorphic to the GIT quotient,
\begin{equation} \label{moduli}
\mathcal{M}_{\kappa,n} \cong \{(\vec{B},\vec{I},\vec{J}) \mid \eqref{Improv1},\eqref{stability}\}/\!\!/ GL(\bK_+|\bK_-)
\end{equation}
\paragraph{}
There is a symmetry \eqref{sa=0}, \eqref{mu=0}, \eqref{BIBJ=0}, and \eqref{JIIJ=0} enjoys, and thus a symmetry of the moduli space \eqref{moduli}: 
We can multiply $B_a$ by a phase $B_a\mapsto q_aB_a$, and compensate with $J_A\mapsto q_AJ_A$, $q_A=q_aq_b$ for $A=(ab)$ as long as the product of $q_a$ is subject to:
\begin{align}\label{product of q}
    \prod_{a=1}^4q_a=1.
\end{align}
If we view $\mathbf{q}={\rm diag}(q_1,q_2,q_3,q_4)$ as diagonal matrix, it belongs to the maximal torus $U(1)^3$ of the group $SU(4)$ rotating $\mathbb{C}^4$.
Hence \eqref{product of q} is the Calabi--Yau four-fold condition.
In the ADHM construction for $\mathbb{R}^4$, we usually consider $SO(4)$ rotation acting on $\mathbb{R}^4$, whose maximal torus $U(1)^2$ give rise to two generic $\Omega$-background parameters for complex momentum map. In the gauge origami, if we start with $SO(8)$ rotation acting on $\mathbb{R}^8=\mathbb{C}^4$ with maximal torus $U(1)^4$, one might expect four generic $\Omega$-background parameters. However  conditions defining the moduli space \eqref{sa=0}, \eqref{BIBJ=0}, and \eqref{JIIJ=0} are real equations, which removes over all $U(1)$ phase rotation \eqref{product of q}, leaving maximal torus $U(1)^3$, which preserves some supersymmetry that act
\begin{align}
    \mathbf{q}\cdot[B_a,I_A,J_A]=[q_aB_a,I_A,q_AJ_A].
\end{align}
As stated, the gauge origami can be viewed as a composition of six copies of ADHM instanton constructions. And doubled to 12 in the case of supergroup. 
Each sub-instanton vector space $\bK_{A,\s}$, $\s=\pm$, has its fixed-points labeled by a set of Young diagrams ${\boldsymbol\lambda}_{A,\s}=(\lambda^{(1)}_{A,\s},\dots,\lambda^{(n_A)}_{A,\s})$, each Young diagram is labeled by a set of non-negative, non-increasing integers $\lambda^{(\alpha)}_{A,\s}=(\lambda_{A,\alpha,1}^\s,\lambda_{A,\alpha,2}^\s,\dots)$,
\begin{equation}
    \lambda_{A,\alpha,1}^\s \geq \lambda_{A,\alpha,2}^\s \geq \cdots \ge 0 ,    
\end{equation}
such that:
\begin{align}
\begin{split}
    & \bK_{A,+}=\bigoplus_{\alpha=1}^{n_A^+}\bK_{A,\alpha,+};\quad \bK_{A,\alpha,+}=\bigoplus_{i=1}^{\ell({\lambda_{A,\alpha}^+)}}\bigoplus_{j=1}^{\lambda_{A,\alpha,i}^+}B_a^{i-1}B_b^{j-1}(I_A e_{A,\alpha}^+); \quad \bN_{A,+}=\mathbb{C}^{n_A^+}=\bigoplus_{\alpha=1}^{n_A^+}\mathbb{C}e_{A,\alpha}^+. \\
    & \bK_{A,-}=\bigoplus_{\alpha=1}^{n_A^-}\bK_{A,\alpha,-};\quad \bK_{A,\alpha,-}=\bigoplus_{i=1}^{\ell({\lambda_{A,\alpha}^-}}\bigoplus_{j=1}^{\lambda_{A,\alpha,i}^-}(B_a^\dagger)^{i-1}(B_b^\dagger)^{j-1}(J_A^\dagger e_{A,\alpha}^-); \quad \bN_{A,-}=\mathbb{C}^{n_A^-}=\bigoplus_{\alpha=1}^{n_A^-}\mathbb{C}e_{A,\alpha}^-.
\end{split}
\end{align}
where $e_{A,\alpha}^\s$ is the fixed basis of the vector space $\mathbb{C}^{n_A^\s}$.
We will also use $\underline{\boldsymbol\lambda}=\{{\boldsymbol\lambda}_A\} = \{ \lambda_{A,\sigma}^{(\alpha)} \}_{\alpha=1,\ldots,n_A^\sigma}^{A \in \underline{6},\sigma=\pm}$ to denote the set of all gauge origami Young diagrams.
\paragraph{}
As in the usual ADHM construction, we denote the character of vector spaces $\bN_A$ and $\bK_A$ as:
\begin{align}
\begin{split}
    & \bN_{A,+} :=\sum_{\alpha=1}^{n_A^+}e^{a_{A,\alpha}^+};\quad
    \bK_{A,+} :=\sum_{\alpha=1}^{n_A^+}e^{a_{A,\alpha}^+}\sum_{(i,j)\in\lambda_{A}^{(\alpha),+}}q_a^{i-1}q_b^{j-1}. \\
    & \bN_{A,-} :=\sum_{\alpha=1}^{n_A^-}e^{a_{A,\alpha}^-};\quad
    \bK_{A,-} :=\sum_{\alpha=1}^{n_A^-}e^{a_{A,\alpha}^-}\sum_{(i,j)\in\lambda_{A}^{(\alpha),-}}q_a^{-i}q_b^{-j}.
\end{split}
\end{align}
Here we are abusing the notation that the vector space and its character use the same notation. 
The character of the tangent bundle of the moduli space defined in eq.~\eqref{moduli} can be written as
\begin{equation}
\CalT_{\underline{\boldsymbol\lambda}} = T_{\underline{\boldsymbol\lambda}} \CalM_{\kappa,n} =  \sum_{A, A'\in\underline{6}} P_{\bar{A}}\bN_A \bK_{A'}^* - P_{123} \sum_{A,A'\in\underline{6}}\bK_A \bK_{A'}^*  - \sum_{A=(12),(13),(14)} q_A \bN_A^*\bN_{\bar{A}},
\end{equation}
with the following notation $a,b,c \in \underline{4}$, $A = (ab) \in \underline{6}$:
\begin{equation}
q_a=e^{\ve_a},\quad P_a=1-q_a,\quad q_A=q_aq_b;\quad P_A=P_aP_b; \quad P_{abc} = P_a P_b P_c.
\end{equation}
Using \eqref{product of q}, it shows $(\ve_a)_{a=1,\ldots,4}$ are subject to the constraint:
\begin{equation}
\sum_{a=1}^4\ve_a=0.
\label{eq:epsilons_constraint}
\end{equation}
Including the one-loop contribution gives the full character of the tangent bundle \cite{NaveenNikita,Nikita:IV}: 
\begin{align}
    \sum_{A \in \underline{6} } \frac{- P_{\min \bar{A} } \bS_A\bS_A^* }{P_A^*} + \sum_{A \cap A'=a} \frac{q_{A'\backslash a} \bS_{A}\bS_{A'}^* + q_{A\backslash a} \bS_{A'} \bS_A^* }{P_a^*} - \sum_{A = (12),(13),(14)} q_{A} \bS_{\bar{A}} \bS_A^* 
\end{align}
with the observable bundle given by the pullback of the inclusion map of the universal bundle associated with $A \in \underline{6}$ over the moduli space,
\begin{align}
    \bS_{A} = \bS_{A,+} - \bS_{A,-}, \quad \bS_{A,\pm} = \bN_{A,\pm} - P_A \bK_{A,\pm}. 
\end{align}


\section{Bethe/Gauge correspondence of eqCM}\label{sec:Bethe}

In this section we aim to establish the gauge theory in correspondence to the eqCM system. 
It is well-known that the classical spectral curve of the eCM integrable system associated with Lie algebra ${\rm Lie}(G)$ can be directly identified with the Seiberg-Witten curve of four dimensional $\CalN=2^*$ theory with gauge group $G$ \cite{Gorsky:1994dj, Nekrasov:1995nq, Martinec:1995by}, see also \cite{DW1, DPLecture}.

There are two gauge theories corresponding to the edCM system: the 4d $\CalN=2^*$ supergroup gauge theory \cite{Chen:2020rxu} and the folded instanton based on gauge origami \cite{Nikita:V,Chen:2019vvt}. 
As the edCM system does not have a proper classical limit, the identification is established through the $qq$-character, a BPS-observabe introduced in \cite{Nikita:I}. It is the double quantum uplift of the Seiberg-Witten curve.

In this section we study the Bethe/Gauge correspondence of the eqCM system. In section~\ref{sec:GO-cons} we compute the partition function of the supergroup folded instanton, whose saddle point equation in the Nekrasov-Shatashvili limit gives the BAE of the eqCM system. 
Finally in section~\ref{sec:orbifold}, an full-type co-dimensional two defect is introduced to the folded instanton. We recover the commuting Hamiltonian of the eqCM system by studying $qq$-character in the presence of the defect. 

\subsection{Gauge origami construction}\label{sec:GO-cons}

We consider the following gauge origami setup: 
\begin{align}\label{def:dd-go}
\begin{split}
    \bN_{12} & = \sum_{\alpha=1}^{N_+} e^{a^+_{\alpha}} - \sum_{\alpha=1}^{N_-} e^{a^-_{\alpha}}, \\
    \bN_{23} & = \sum_{\beta=1}^{M_+} e^{b^+_{\beta}} - \sum_{\beta=1}^{M_-} e^{b^-_{\beta}}.
\end{split}
\end{align}
with four sets of Coulomb moduli $\ba^\pm = (a_1^\pm,\dots,a_{N_\pm}^\pm)$, $\bb^\pm = (b_1^\pm,\dots,b_{M_\pm}^\pm)$. 
Namely, $n_{12}^\pm = N_\pm$, $n_{23}^\pm = M_\pm$, and $n_A^\pm = 0$ for $A \in \underline{6} \backslash \{ (12), (23) \}$.
The gauge origami with two sets of D-branes sharing a two-dimsnional subspace is called \emph{folded instanton}. 
The folded instanton build based on supergroups are characterized by four tuples of Young diagrams $\boldsymbol\lambda_{12,\pm}=(\lambda_{12,\pm}^{(1)},\dots,\lambda_{12,\pm}^{(N_\pm)})$ and $\boldsymbol\lambda_{23,\pm} = (\lambda_{23,\pm}^{(1)},\dots,\lambda_{23,\pm}^{(M_\pm)})$.  
The partition function is given by
\begin{align}\label{def:Partition}
\begin{split}
    & \CalZ_S (\ba^\pm,\bb^\pm,\ve_1,\ve_2,\ve_3) \\
    & = \sum_{{\boldsymbol\lambda_{12,\pm}},\boldsymbol\lambda_{23,\pm}} \frac{\kq^{|\boldsymbol\lambda_{12,+}|+|\boldsymbol\lambda_{23,+}|}}{\kq^{|\boldsymbol\lambda_{12,-}|+|\boldsymbol\lambda_{23,-}|}} \BE \left[ -\frac{P_3\bS_{12}\bS_{12}^*}{P_{12}^*} - \frac{P_1\bS_{23}\bS_{23}^*}{P_{23}^*} + \frac{q_3\bS_{12}\bS_{23}^* + q_1\bS_{23}\bS_{12}^*}{P_2^*} \right] \\
    & = \sum_{{\boldsymbol\lambda_{12,\pm}},\boldsymbol\lambda_{23,\pm}} \kq^{|\boldsymbol\lambda_{12,+}|+|\boldsymbol\lambda_{23,+}|-|\boldsymbol\lambda_{12,-}|-|\boldsymbol\lambda_{23,-}|} \CalZ_{S}[{\boldsymbol\lambda_{12,\pm}},\boldsymbol\lambda_{23,\pm}]
\end{split}
\end{align}
where we have defined Chern characters 
\begin{align}
\begin{split}
    & \bN_{12,\pm} = \sum_{\alpha=1}^{N_\pm} e^{a^\pm_{\alpha}}, \quad \bK_{12,\pm} = \sum_{\alpha=1}^{N_\pm} \sum_{(i,j)\in \lambda_{12,\pm}^{(\alpha)} } e^{a^\pm_{\alpha} + \frac{\ve_+}{2}} q_1^{\pm\left(i-\frac{1}{2}\right)}q_2^{\pm\left(j-\frac{1}{2}\right)} := \sum_{\Box \in \lambda_{12,\pm} } e^{c_\Box^{12,\pm} } \\
    & \bN_{23,\pm} = \sum_{\beta=1}^{M_\pm} e^{b^\pm_{\beta}}, \quad \bK_{23,\pm} = \sum_{\beta=1}^{M_\pm} \sum_{(i,j)\in \lambda_{23,\pm}^{(\beta)}} e^{b^\pm_{\beta} - \frac{\ve_+}{2}} q_3^{\pm\left(i-\frac{1}{2}\right)}q_2^{\pm\left(j-\frac{1}{2}\right)} := \sum_{\Box \in \lambda_{32,\pm} } e^{c_\Box^{32,\pm} }
\end{split}
\end{align}
with $\ve_+ = \ve_1+\ve_2$
and the observable bundle,
\begin{align}
\begin{split}
    & \bS_{12} = \bS_{12,+} - \bS_{12,-}, \quad \bS_{12,\pm} = \bN_{12,\pm} - P_1P_2 \bK_{12,\pm}, \\
    & \bS_{23} = \bS_{23,+} - \bS_{23,-}, \quad \bS_{23,\pm} = \bN_{23,\pm} - P_2P_3 \bK_{23,\pm}.
\end{split}
\end{align}
Given a character $\bX = \sum_{\ti} \tn_\ti e^{\tx_\ti}$, we denote its dual character $\bX^* = \sum_{\ti} \tn_\ti e^{-\tx_\ti}$. $\BE$ is the index functor that converts the additive Chern characters to the multiplicative class (equivariant Euler class),
\begin{align}\label{def:E-operator}
    \BE \left[ \sum_{\ti}\tn_\ti e^{\tx_\ti} \right] = \prod_\ti \tx_\ti^{-\tn_\ti}. 
\end{align}

\subsection{Bethe ansatz equation} \label{sec:BAE}
In the Nekrasov-Shatashivilli limit (NS-limit for short) $\ve_2 \to 0$, the partition function \eqref{def:Partition} has the following asymptotics,
\begin{align}
    \ve_2 \log \CalZ_S(\ba^\pm,\bb^\pm,\ve_1,\ve_2,\ve_3) = -\CalW(\ba^\pm,\bb^\pm,\ve_1,\ve_3) + \CalO(\ve_2)
    .
\end{align}
Here $\CalW(\ba^\pm,\bb^\pm,\ve_1,\ve_3)$ is the effective twisted superpotential of the two-dimensional degrees of freedom on $\mathbb{C}_2$ \cite{Nikita-Pestun-Shatashvili,Nikita-Shatashvili}. 
We say that the partition function in the NS-limit is dominated by the \emph{limit shape} instanton configuration $(\boldsymbol\Lambda_{12,\pm}$, $\boldsymbol\Lambda_{23,\pm})$ defined by
\begin{align}
    \CalZ_S[\boldsymbol\Lambda_{12,\pm},\boldsymbol\Lambda_{23,\pm}] = e^{-\frac{\CalW(\ba^\pm,\bb^\pm,\ve_1,\ve_3)}{\ve_2}}. 
\end{align}
We denote
\begin{align}
\begin{split}
    & \bS_{12,\pm}[\boldsymbol\Lambda_{12,\pm}] = P_1\bX_{12,\pm}, \ \bX_{12,\pm} = \sum_{\alpha=1}^{N_\pm} \sum_{i=1}^\infty e^{x_{\alpha i}^{12,\pm} },  \\
    & \bS_{23,\pm}[\boldsymbol\Lambda_{23,\pm}] = P_1 \bX_{23,\pm}, \ \bX_{23,\pm} = \sum_{\beta=1}^{M_\pm} \sum_{i=1}^\infty e^{x_{\beta i}^{23,\pm} } ,
\end{split} 
\end{align}
with
\begin{align}
\begin{split}
    & x_{\alpha i}^{1,+} = a_\alpha^+ + (i-1)\ve_1 + \ve_2 \Lambda^{(\alpha)}_{12,+,i}, 
    \\
    & x_{\alpha\ri}^{1,-} = a_\alpha^- - i\ve_1 - \ve_2 \Lambda^{(\alpha)}_{12,-,i}, 
    \\
    & x_{\beta i}^{3,+} = b_\beta^+ + (i-1)\ve_3 + \ve_2 \Lambda^{(\beta)}_{23,+,i}, 
    \\
    & x_{\beta i}^{3,-} = b_\beta^- -i\ve_3 - \ve_2 \Lambda^{(\beta)}_{23,-,i}. 
\end{split}
\end{align}

The saddle point of the instanton pseudo-measure in \eqref{def:Partition} can be obtained by considering small variation on top of the limit shape. This can be realised by adding an additional instanton in the limit shape configuration $(\boldsymbol\Lambda_{12,\pm},\boldsymbol\Lambda_{23,\pm})$. The potential position for a new box to be added is exactly at $x_{\alpha i}^{1,+}$, $x_{\alpha i}^{1,-}$, $x_{\beta i}^{3,+}$ or $x_{\beta i}^{3,-}$. Let us first consider adding an instanton in $\bS_{12,+}$ at $e^{x_{\alpha i}^{1,+}}$. Then, the saddle point equation reads
\begin{align}
\begin{split}
    1 & = \kq \frac{\CalZ_S[\boldsymbol\Lambda_{12,\pm}+\Box, \boldsymbol\Lambda_{23,\pm}]}{\CalZ_S[\boldsymbol\Lambda_{12,\pm}, \boldsymbol\Lambda_{23,\pm}]} \\
    & = \kq \BE \left[ q_{12} P_3 e^{x_{\alpha i}^{1,+}} (\bS_{12}-e^{x_{\alpha i}^{1,+}})^* + P_3^* e^{x_{\alpha i}^{1,+}} \bS_{12}^* + (q_3q_2 P_1 + P_1^*) e^{x_{\alpha i}^{1,+}} \bS_{23}^* \right] \\
    & = \kq \frac{Y_1(x_{\alpha i}^{1,+}-\ve_3) Y_1(x_{\alpha i}^{1,+}-\ve_4)}{Y_1(x_{\alpha i}^{1,+}+\ve_1)Y_1(x_{\alpha i}^{1,+})} \frac{Y_3(x_{\alpha i}^{1,+}-\ve_1)Y_3(x_{\alpha i}^{1,+}-\ve_4)}{Y_3(x_{\alpha i}^{1,+}+\ve_3)Y_3(x_{\alpha i}^{1,+})} \\
    & = \kq \frac{Q(x_{\alpha i}^{1,+}-\ve_1)Q(x_{\alpha i}^{1,+}-\ve_3)Q(x_{\alpha i}^{1,+}-\ve_4)}{Q(x_{\alpha i}^{1,+}+\ve_1)Q(x_{\alpha i}^{1,+}+\ve_3)Q(x_{\alpha i}^{1,+}+\ve_4)} .
\end{split}
\end{align}
Here we define
\begin{align}
    Y_1(x) = \BE \left[ -e^x \bS_{12}^* \right] = \frac{Q_1(x)}{Q_1(x-\ve_1)}, \quad Y_3(x) = \BE \left[ - e^x \bS_{23}^* \right] = \frac{Q_3(x)}{Q_3(x-\ve_3)}.
\end{align}
with
\begin{subequations}\label{def:Q}
\begin{gather}
    Q_1(x) = \frac{ \prod\limits_{\alpha=1}^{N_+} \prod\limits_{i=1}^\infty (x-x_{\alpha i}^{1,+} )}{\prod\limits_{\alpha=1}^{N_-} \prod\limits_{i=1}^\infty (x-x_{\alpha i}^{1,-}) }, \qquad Q_3(x) = \frac{ \prod\limits_{\beta=1}^{M_+} \prod\limits_{i=1}^\infty (x-x_{\beta i}^{3,+} )}{\prod\limits_{\beta=1}^{M_-} \prod\limits_{i=1}^\infty (x-x_{\beta i}^{3,-}) }, \\
    Q(x) = Q_1(x)Q_3(x) = \frac{ \prod\limits_{\alpha=1}^{N_+} \prod\limits_{i=1}^\infty (x-x_{\alpha i}^{1,+} )}{\prod\limits_{\alpha=1}^{N_-} \prod\limits_{i=1}^\infty (x-x_{\alpha i}^{1,-}) } \frac{ \prod\limits_{\beta=1}^{M_+} \prod\limits_{i=1}^\infty (x-x_{\beta i}^{3,+} )}{\prod\limits_{\beta=1}^{M_-} \prod\limits_{i=1}^\infty (x-x_{\beta i}^{3,-}) } .
\end{gather}    
\end{subequations}
Similar calculation can be done for the other three cases. We have the following Bethe ansatz equation (BAE):
\begin{align}\label{eq:BAE}
    1 = \kq \frac{Q(x-\ve_1)Q(x-\ve_3)Q(x-\ve_4)}{Q(x+\ve_1)Q(x+\ve_3)Q(x+\ve_4)}, \quad x = x_{\alpha i}^{1,\pm}, \ x_{\beta i}^{3,\pm}.
\end{align}
In the $\ve_2 \to 0$ limit, the constraint \eqref{eq:epsilons_constraint} becomes $\ve_1+\ve_3+\ve_4 = 0$. 
The eCM, edCM and eqCM all share the same from of the Bethe equation \eqref{eq:BAE} with different forms of $Q(x)$~\cite{Nikita-Shatashvili,Chen:2012we,Litvinov:2013zda,Feigin:2015raa,FJMM,Chen:2019vvt,Chen:2020rxu,Prochazka:2023zdb}. 
The same BAE is also observed in the $\mathbb{C}^3$ system \cite{Cao:2023lon,Kimura:2023bxy} with a different $Q$-function. 

\subsection{\textit{qq}-character}\label{sec:qq}

A class of BPS observables called $qq$-character is introduced in \cite{Nikita:I}, which are the quantum uplift of the $q$-character \cite{Nikita-Pestun-Shatashvili}, the double-quantum uplift of the Seiberg-Witten curve. 
The main statement in \cite{Nikita:II} proves certain vanishing theorem for the expectation value of the $qq$-character. These vanishing equations, called \emph{non-perturbative Dyson-Schwinger equation}, can be used to derive KZ-type equations \cite{kz,jeong2021intersecting,Jeong:2023qdr}. 
In the NS-limit, these KZ-type equations becomes Schr\"{o}dinger-type equation, producing the quantum Hamiltonian of the corresponding integrable system \cite{Nekrasov:2017gzb,Lee:2020hfu}.

The $qq$-character of the folded instanton with supergroup \eqref{def:dd-go} can be constructed by adding one additional D-brane in the $\bN_{34}$ and $\bN_{14}$ respectively. We consider adding a single D-brane in $\bN_{l4,+} = e^x$, $l=1,3$. The instanton configuration in the $\bN_{l4}$  space is labeled by a single Young diagram $\lambda_{l4,+}$
\begin{align}
    \bK_{l4,+} = \sum_{(\bi,\bj)\in \lambda_{l4,+}} e^x q_l^{\bi-1} q_4^{\bj-1}, \quad \bS_{l4} = \bN_{l4,+} - P_lP_4\bK_{l4,+}, \ l=1,3.
\end{align}
The spiked instanton partition function is given by
\begin{align}
\begin{split}
    \CalZ_X = &  \sum_{{\boldsymbol\lambda_{12,\pm}},\boldsymbol\lambda_{23,\pm},\lambda_{34,+}} \frac{\kq^{|\boldsymbol\lambda_{12,+}|+|\boldsymbol\lambda_{23,+}|}}{\kq^{|\boldsymbol\lambda_{12,-}|+|\boldsymbol\lambda_{23,-}|}} \kq^{|\lambda_{l4,+}|} \\
    & \qquad \BE \left[ -\frac{P_3\bS_{12}\bS_{12}^*}{P_{12}^*} - \frac{P_1\bS_{23}\bS_{23}^*}{P_{23}^*} + \frac{q_3\bS_{12}\bS_{23}^* + q_1\bS_{23}\bS_{12}^*}{P_2^*} \right. \\
    & \qquad \qquad \left. - \frac{P_{4-l}\bS_{l4}\bS_{l4}^*}{P_{l4}^*} - q_{4-l}q_{2} \bS_{l4}\bS_{(4-l)2}^* + \frac{q_2P_{4-l}\bS_{l4}\bS_{l2}^*}{P_l^*} \right] \\
    = & \sum_{{\boldsymbol\lambda_{12,\pm}},\boldsymbol\lambda_{23,\pm}} \frac{\kq^{|\boldsymbol\lambda_{12,+}|+|\boldsymbol\lambda_{23,+}|}}{\kq^{|\boldsymbol\lambda_{12,-}|+|\boldsymbol\lambda_{23,-}|}} \CalZ_{S}[{\boldsymbol\lambda_{12,\pm}},\boldsymbol\lambda_{23,\pm}] \CalX_l(x)[{\boldsymbol\lambda_{12,\pm}},\boldsymbol\lambda_{23,\pm}]
\end{split}
\end{align}
The $qq$-character $\CalX_1(x)$ (resp. $\CalX_3$) is obtained by integrating over the degree of freedom in the $\mathbb{C}_3 \times \mathbb{C}_4$ ($\mathbb{C}_1 \times \mathbb{C}_4$) space, which becomes an ensemble over the instanton configuration in the context of localization,
\begin{subequations}
\begin{align}
    \CalX_1(x) = & \ Y_1(x+\ve_{12}) \frac{Q_3(x+\ve_{12})}{Q_3(x+\ve_2)} \sum_{\lambda_{34,+}} \kq^{|\lambda_{34,+}|} B_{12}[\lambda_{34,+}] \\
    & \times \prod_{(\bi,\bj)\in\lambda_{34,+}} \frac{Y_1(x+s_{1,\bi\bj}-\ve_3)Y_1(x+s_{1,\bi\bj}-\ve_4)}{Y_1(x+s_{1,\bi\bj})Y_1(x+s_{1,\bi\bj}-\ve_{34})} \frac{Y_3(x+s_{1,\bi\bj}-\ve_1)Y_3(x+s_{1,\bi\bj}-\ve_4)}{Y_3(x+s_{1,\bi\bj})Y_3(x+s_{1,\bi\bj}-\ve_{14})} \nonumber\\
    \CalX_3(x) = & \ Y_3(x+\ve_{23}) \frac{Q_1(x+\ve_{23})}{Q_1(x+\ve_2)} \sum_{\lambda_{14,+}} \kq^{|\lambda_{14,+}|} B_{23}[\lambda_{14,+}] \\
    & \times \prod_{(\bi,\bj)\in\lambda_{14,+}} \frac{Y_3(x+s_{3,\bi\bj}-\ve_1)Y_3(x+s_{3,\bi\bj}-\ve_4)}{Y_3(x+s_{3,\bi\bj})Y_3(x+s_{3,\bi\bj}-\ve_{14})} \frac{Y_1(x+s_{3,\bi\bj}-\ve_3)Y_1(x+s_{3,\bi\bj}-\ve_4)}{Y_1(x+s_{3,\bi\bj})Y_1(x+s_{3,\bi\bj}-\ve_{34})} \nonumber
\end{align}
\end{subequations}
where $s_{1,\bi\bj} = (\bi-1)\ve_3+(\bj-1)\ve_4$, $s_{3,\bi\bj} = (\bi-1)\ve_1+(\bj-1)\ve_4$, $\ve_{ab}=\ve_a+\ve_b$. We also define:
\begin{subequations}
\begin{align}
    & B_{12}[\lambda_{34,+}] = \prod_{(\bi,\bj)\in \lambda_{34,+}} \left[ 1 + \frac{\ve_1\ve_2}{(\ve_3(l_{\bi\bj}+1) - \ve_4a_{\bi\bj})(\ve_3(l_{\bi\bj}+1) - \ve_4a_{\bi\bj}+\ve_{12})} \right] \\
    & B_{23}[\lambda_{14,+}] = \prod_{(\bi,\bj)\in \lambda_{14,+}} \left[ 1 + \frac{\ve_3\ve_2}{(\ve_1(l_{\bi\bj}+1) - \ve_4a_{\bi\bj})(\ve_1(l_{\bi\bj}+1) - \ve_4a_{\bi\bj}+\ve_{12})} \right]
\end{align}
\end{subequations}
Let us explain the notation here: Given a Young diagram $\lambda=(\lambda_1,\dots,\lambda_{\ell(\lambda)})$, we denote $a_{\bi\bj} = \lambda_\bi-\bj$ the "arm" of the associated box $(\bi,\bj)\in \lambda$, and $l_{\bi\bj} = \lambda^\text{T}_\bj-\bi$ the "leg" associated to the same box. 
Notice that both $B_{12}$ and $B_{23}$ approaches to 1 in the Nekrasov-Shatashvili limit $\ve_2 \to 0$. 
The BAE \eqref{eq:BAE} can be obtained from the $qq$-character in the Nekrasov-Shatashvili limit. We rewrite the $qq$-character in $Q(x)$ \eqref{def:Q} in $\ve_2\to 0$ limit:
\begin{subequations}
\begin{align}
    \CalX_1(x) = & \ \frac{Q(x+\ve_{1})}{Q(x)} \sum_{\lambda_{34,+}} \kq^{|\lambda_{34,+}|} \prod_{(\bi,\bj)\in\lambda_{34,+}} \frac{Q(x+s_{1,\bi\bj}-\ve_1)Q(x+s_{1,\bi\bj}-\ve_3)Q(x+s_{1,\bi\bj}-\ve_4)}{Q(x+s_{1,\bi\bj}+\ve_1)Q(x+s_{1,\bi\bj}+\ve_3)Q(x+s_{1,\bi\bj}+\ve_4)} \\
    \CalX_3(x) = & \frac{Q(x+\ve_{3})}{Q(x)} \sum_{\lambda_{14,+}} \kq^{|\lambda_{14,+}|} \frac{Q(x+s_{1,\bi\bj}-\ve_1)Q(x+s_{1,\bi\bj}-\ve_3)Q(x+s_{1,\bi\bj}-\ve_4)}{Q(x+s_{1,\bi\bj}+\ve_1)Q(x+s_{1,\bi\bj}+\ve_3)Q(x+s_{1,\bi\bj}+\ve_4)} 
\end{align}
\end{subequations}


\subsection{Orbifold surface defect} \label{sec:orbifold}

We now introduce $\BZ_{L'}$ orbifold on the geometry $\BC^4$ induced by the discrete action $\BZ_{L'} : (\bz_1,\bz_2,\bz_3,\bz_4) \mapsto (\bz_1, \eta \bz_2, \bz_3, \eta^{-1} \bz_4) $, where $\eta = e^{\frac{2\pi \ri}{L'} }$ is the $L'$-th root of unity. 
The orbifold defect is characterized by a coloring function 
$$
    c:\{\alpha_+\}_{\alpha_+=1}^{N_+} \cup \{\alpha_-\}_{\alpha_-=1}^{N_-} \cup \{\beta_+\}_{\beta_+=1}^{M_+} \cup \{\beta_-\}_{\beta_-=1}^{M_-} \to \BZ_L
$$
that assigns the Coulomb moduli parameters $a_{\pm,\alpha}$, $b_{\pm,\beta}$ to the $\o$-th one-dimensional irreducible representation $\CalR_\o$ of $\BZ_L$. The defect induced by this orbifold is called \emph{regular/full-type surface defect} if the following are true: 
\begin{itemize}
    \item $L' = L = N_++N_-+M_++M_-$,
    \item $c$ is bijective. 
\end{itemize}
From now on we will consider only the regular/full-type surface defect with the following coloring function:
\begin{align}
\begin{split}
    & c(\alpha_+) = \alpha_+ - 1, \quad 
    c(\alpha_-) = \alpha_- + N_+ - 1, \\ 
    & c(\beta_+) = \beta_+ + N_+ + N_- - 1, \quad
    c(\beta_-) = \beta_- + N_+ + N_- + M_+ - 1 .
\end{split}
\end{align}
For this coloring function we will denote
\begin{align}
\begin{split}
    & [\alpha_+] = \{0,\dots,N_+-1\}, \\
    & [\alpha_-] = \{N_+,\dots,N_++N_--1\}, \\
    & [\beta_+] = \{N_++N_-,\dots,N_++N_-+M_+-1\}, \\ 
    & [\beta_-] = \{N_++N_-+M_+,\dots,N_++N_-+M_++M_--1 \}.
\end{split}
\end{align}
such that $[\alpha_+] \cup [\alpha_-] \cup [\beta_+] \cup [\beta_-] = \{0,\dots,N_++N_-+M_++M_--1\}$, which is the range of the index $\o$. The orbifold also splits the instanton counting parameter $\kq$ into $L=N_++N_-+M_++M_-$ fractional counting parameters $\kq_\o$ for each representation $\CalR_\o$ of $\BZ_L$. The fractional couplings $\{\kq_\o\}$ are related to the bulk instanton parameter by
\begin{align}
    \kq = \prod_{\o=0} ^{L-1} \kq_\o^{p(\o)}, \quad \kq_\o^{p(\o)} = \frac{\fz_{\o}}{\fz_{\o-1}}
\end{align}
with the parity function $p(\o)$ defined by
\begin{align}
    p(\o) = 
    \begin{cases}
        1, & \o \in [\alpha_+] \cup [\beta_+] \\
        -1 & \o \in [\alpha_-] \cup [\beta_-]
    \end{cases}. 
\end{align}
We also define the particle coordinates:
\begin{align}
    \fz_\o = 
    \begin{cases}
        e^{\rx_\alpha^+} & \o \in [\alpha_+], \ c(\alpha) = \o \\
        e^{\rx_\alpha^-} & \o \in [\alpha_-], \ c(\alpha) = \o \\
        e^{\ry_\beta^+} & \o \in [\beta_+], \ c(\beta) = \o \\
        e^{\ry_\beta^-} & \o \in [\beta_-], \ c(\beta) = \o
    \end{cases}.
\end{align}

Fractional $qq$-characters are: 
\begin{subequations}
\begin{align}
    \CalX_{1,\o}(x) = & \ Y_{1,\o+1}(x+\ve_{12}) \frac{Q_{3,\o+1}(x+\ve_{12})}{Q_{3,\o+1}(x+\ve_2)} \sum_{\lambda_{34,+}} \BB_{12,\o}[\lambda_{34,+}] \\
    & \times \prod_{(\bi,\bj)\in\lambda_{34,+}} \frac{Y_{1,\o+1-\bj}(x+s_{1,\bi\bj}-\ve_3)Y_{1,\o+2-\bj}(x+s_{1,\bi\bj}-\ve_4)}{Y_{1,\o+1-\bj}(x+s_{1,\bi\bj})Y_{1,\o+2-\bj}(x+s_{1,\bi\bj}-\ve_{34})} \nonumber \\
    & \qquad \qquad  \times \frac{Y_{3,\o+1-\bj}(x+s_{1,\bi\bj}-\ve_1)Y_{3,\o+2-\bj}(x+s_{1,\bi\bj}-\ve_4)}{Y_{3,\o+1-\bj}(x+s_{1,\bi\bj})Y_{3,\o+2-\bj}(x+s_{1,\bi\bj}-\ve_{14})} \nonumber\\
    \CalX_{3,\o}(x) = & \ Y_{3,\o+1}(x+\ve_{23}) \frac{Q_{1,\o+1}(x+\ve_{23})}{Q_{1,\o+1}(x+\ve_2)} \sum_{\lambda_{14,+}} \BB_{23,\o}[\lambda_{14,+}] \\
    & \times \prod_{(\bi,\bj)\in\lambda_{14,+}} \frac{Y_{3,\o+1-\bj}(x+s_{3,\bi\bj}-\ve_1)Y_{3,\o+2-\bj}(x+s_{3,\bi\bj}-\ve_4)}{Y_{3,\o+1-\bj}(x+s_{3,\bi\bj})Y_{3,\o+2-\bj}(x+s_{3,\bi\bj}-\ve_{14})} \nonumber\\
    & \qquad \qquad \times \frac{Y_{1,\o+1-\bj}(x+s_{3,\bi\bj}-\ve_3)Y_{1,\o+2-\bj}(x+s_{3,\bi\bj}-\ve_4)}{Y_{1,\o+1-\bj}(x+s_{3,\bi\bj})Y_{1,\o+2-\bj}(x+s_{3,\bi\bj}-\ve_{34})} \nonumber
\end{align}
\end{subequations}
The fractional $Y$-functions are defined by
\begin{subequations}
\begin{align}
    Y_{1,\o}(x) = & \frac{ \prod\limits_{\alpha_+=1}^{N_+} (x-a^+_{\alpha_+})^{\d(c(\alpha_+)-\o)} }{ \prod\limits_{\alpha_-=1}^{N_-} (x-a^-_{\alpha_-})^{\d(c(\alpha_-)-\o)} } \prod_{\Box \in \EK^{12,+}_\o } \frac{x-c_\Box^{12,+} - \ve_1 }{x-c_\Box^{12,+}} \prod_{\Box\in \EK_{\o+1}^{12,+} } \frac{x-c_\Box^{12,+}-\ve_2}{x-c_\Box^{12,+}-\ve_{12}} \nonumber\\
    & \qquad \times \prod_{\Box \in \EK^{12,+}_\o } \frac{x-c_\Box^{12,-}}{x-c_\Box^{12,-} - \ve_1 } \prod_{\Box\in \EK_{\o+1}^{12,-} } \frac{x-c_\Box^{12,-}-\ve_{12}}{x-c_\Box^{12,-}-\ve_2} \\
    Y_{1,\o}(x) = & \frac{ \prod\limits_{\beta_+=1}^{M_+} (x-b^+_{\beta_+})^{\d(c(\beta_+)-\o)} }{ \prod\limits_{\beta_-=1}^{M_-} (x-b^-_{\beta_-})^{\d(c(\beta_-)-\o)} } \prod_{\Box \in \EK^{32,+}_\o } \frac{x-c_\Box^{32,+} - \ve_3 }{x-c_\Box^{32,+}} \prod_{\Box\in \EK_{\o+1}^{32,+} } \frac{x-c_\Box^{32,+}-\ve_2}{x-c_\Box^{32,+}-\ve_{32}} \nonumber\\
    & \qquad \times \prod_{\Box \in \EK^{32,+}_\o } \frac{x-c_\Box^{32,-}}{x-c_\Box^{32,-} - \ve_1 } \prod_{\Box\in \EK_{\o+1}^{12,-} } \frac{x-c_\Box^{32,-}-\ve_{32}}{x-c_\Box^{32,-}-\ve_2}
\end{align}
\end{subequations}
The boxes in the Young diagrams are colored in the presence of orbifold: 
\begin{subequations}\label{def:frac-EK}
\begin{align}
    \EK_\o^{12,+} & = \{ (\alpha_+,(i,j))\ |\ \alpha_+ =1,\dots,N_+, \ (i,j) \in \lambda_{12,+}^{(\alpha_+)}, \ c(\alpha) + j-1 \equiv \o \text{ mod } L \}; \\
    \EK_\o^{12,-} & = \{ (\alpha_-,(i,j))\ |\ \alpha_- =1,\dots,N_-, \ (i,j) \in \lambda_{12,-}^{(\alpha_-)}, \ c(\alpha) - j \equiv \o \text{ mod } L  \}; \\
    \EK_\o^{32,+} & = \{ (\beta_+,(i,j))\ |\ \beta_+ =1,\dots,M_+, \ (i,j) \in \lambda_{32,+}^{(\beta_+)}, \ c(\beta) + j-1 \equiv \o \text{ mod } L \}; \\
    \EK_\o^{32,-} & = \{ (\beta_-,(i,j))\ |\ \beta_- =1,\dots,M_-, \ (i,j) \in \lambda_{32,-}^{(\beta_-)}, \ c(\beta) -j \equiv \o \text{ mod } L \}.
\end{align}
\end{subequations}

Let us address the $\BB_{12,\o}[\lambda_{34,+}]$ and $\BB_{23,\o}[\lambda_{14,+}]$ here. They are the orbifolded versions of $\kq^{|\lambda_{34,+}|}B_{12}[\lambda_{34,+}]$ and $\kq^{|\lambda_{14,+}|}B_{23}[\lambda_{14,+}]$ respectively. Here we change the stability condition for those $p(\o)=1$. Each is given by
\begin{subequations}
\begin{align}
    & \BB_{12,\o}[\lambda_{34,+}] = \left. \prod_{(\bi,\bj)\in \lambda_{34,+}} \kq_{\o+1-\bj} B_{1,\o} ( (-1)^{\tp(\o)} \ve_3 (l_{\bi\bj}+1) ) \right|_{a_{\bi\bj}=0} ,  \\
    & \BB_{23,\o}[\lambda_{14,+}] = \left. \prod_{(\bi,\bj)\in \lambda_{14,+}} \kq_{\o+1-\bj} B_{3,\o} ( (-1)^{\tp(\o)} \ve_1 (l_{\bi\bj}+1) ) \right|_{a_{\bi\bj}=0}, \ 
\end{align}
\end{subequations}
with
\begin{subequations}
\begin{align}
    & B_{1,\o}(x) = 1 + \frac{\ve_1}{x} , \\
    & B_{3,\o}(x) = 1 + \frac{\ve_3}{x} .
\end{align}
\end{subequations}

One way to think about this configuration is that the orbifolding now splits the instanton partition function into $L=N_++N_-+M_++M_-$ copies of $U(1)$ sub-partitions. Each elements in $\bK_\o$ is counted by fractional coupling $\kq_\o$ instead of the original $\kq$. To evaluate the ensemble over all possible Young diagrams, let us introduce a new representation for a Young diagram $\lambda$:
\begin{align}
    \lambda = (1^{\ell_0}2^{\ell_{1}}\cdots (L-1)^{\ell_{L-2}} L^{\boldsymbol\ell}). 
\end{align}
Each $\ell_{r-1} = \sum\limits_{J=0}^\infty \ell_{r-1,J}$, where $\ell_{r-1,J} = \lambda^\text{T}_{r+LJ} - \lambda^\text{T}_{r+1+LJ} $ is the difference between the number of boxes in of two neighboring columns, $r=1,\dots,L$. The last one $\boldsymbol\ell = \sum\limits_{J=1}^\infty \lambda^\text{T}_{LJ}$ counts how many times a full combination $\kq$ shows up. The summation over the Young diagrams gives
\begin{subequations}
\begin{align}
\begin{split}
    \BB_{12,\o}(\vec{z};\tau)
    & = \sum_{\lambda} \BB_{12,\o}[\lambda] = \sum_{\ell_0,\dots,\ell_{N-1},{\boldsymbol\ell}} \prod_{\o'=0}^{L-1} \left[ \prod_{j=1}^{\ell_{\o'}} \frac{\left( (-1)^{\tp(\o)} j + \frac{\ve_1}{\ve_3} \right)}{(-1)^{\tp(\o)} j} \right] \left( \frac{\fz_\o}{\fz_{\o'}} \right)^{\ell_{\o'}} \kq^{\boldsymbol\ell} \\
    & = 
    \begin{cases}
        \prod\limits_{\o'<\o} \left[ \frac{1}{(\frac{\fz_\o}{\fz_{\o'}};\kq)_\infty} \right]^{\frac{\ve_1+\ve_3}{\ve_3}} \prod\limits_{\o'\geq\o} \left[ \frac{1}{(\frac{\fz_\o}{\fz_{\o'}}\kq;\kq)_\infty} \right]^{\frac{\ve_1+\ve_3}{\ve_3}}  & (\tp(\o) = 0) \\
        \prod\limits_{\o'<\o} \left[ {(\frac{\fz_\o}{\fz_{\o'}};\kq)_\infty} \right]^{\frac{\ve_1}{\ve_3}} \prod\limits_{\o'\geq\o} \left[ {(\frac{\fz_\o}{\fz_{\o'}}\kq;\kq)_\infty} \right]^{\frac{\ve_1}{\ve_3}} & (\tp(\o) = 1)
    \end{cases}
\end{split}
\end{align}
\begin{align}
\begin{split}
    \BB_{32,\o}(\vec{z};\tau)
    & = \sum_{\lambda} \BB_{32,\o}[\lambda] = \sum_{\ell_0,\dots,\ell_{N-1},{\boldsymbol\ell}} \prod_{\o'=0}^{L-1} \left[ \prod_{j=1}^{\ell_{\o'}} \frac{\left( (-1)^{\tp(\o)} j + \frac{\ve_1}{\ve_3} \right)}{(-1)^{\tp(\o)} j} \right] \left( \frac{\fz_\o}{\fz_{\o'}} \right)^{\ell_{\o'}} \kq^{\boldsymbol\ell} \\
    & = 
    \begin{cases}
        \prod\limits_{\o'<\o} \left[ \frac{1}{(\frac{\fz_\o}{\fz_{\o'}};\kq)_\infty} \right]^{\frac{\ve_3+\ve_1}{\ve_1}} \prod\limits_{\o'\geq\o} \left[ \frac{1}{(\frac{\fz_\o}{\fz_{\o'}}\kq;\kq)_\infty} \right]^{\frac{\ve_3+\ve_1}{\ve_1}} & (\tp(\o)= 0) \\
        \prod\limits_{\o'<\o} \left[ {(\frac{\fz_\o}{\fz_{\o'}};\kq)_\infty} \right]^{\frac{\ve_3}{\ve_1}} \prod\limits_{\o'\geq\o} \left[ {(\frac{\fz_\o}{\fz_{\o'}}\kq;\kq)_\infty} \right]^{\frac{\ve_3}{\ve_1}} & (\tp(\o)= 1)
    \end{cases}
\end{split}
\end{align}
\end{subequations}

Performing large $x$ expansion of $Y_{1,\o}$ and $Y_{3,\o}$ under the orbifold, we have
\begin{subequations}
\begin{align}
    Y_{1,\o}(x) & = (x-a^+_{c^{-1}(\o)}) \exp \left[ \frac{\ve_1}{x} \nu_{\o-1}^{12} + \frac{\ve_1\ve_2}{x^2} k_{\o-1}^{12} + \frac{\ve_1}{x^2} D_{\o-1}^{12} + \CalO(x^{-3})  \right], \quad \o \in [\alpha_+], \\
    Y_{1,\o}(x) & = \frac{1}{(x-a^-_{c^{-1}(\o)} )} \exp \left[ \frac{\ve_1}{x} \nu_{\o-1}^{12} + \frac{\ve_1\ve_2}{x^2} k_{\o-1}^{12} + \frac{\ve_1}{x^2} D_{\o-1}^{12} + \CalO(x^{-3})  \right], \quad \o \in [\alpha_-], \\
    Y_{3,\o}(x) & = (x-b^+_{c^{-1}(\o)}) \exp \left[ \frac{\ve_3}{x} \nu_{\o-1}^{32} + \frac{\ve_3\ve_2}{x^2} k_{\o-1}^{32} + \frac{\ve_3}{x^2} D_{\o-1}^{32} + \CalO(x^{-3})  \right], \quad \o \in [\beta_+], \\
    Y_{3,\o}(x) & = \frac{1}{(x-b^-_{c^{-1}(\o)} )} \exp \left[ \frac{\ve_3}{x} \nu_{\o-1}^{32} + \frac{\ve_3\ve_2}{x^2} k_{\o-1}^{32} + \frac{\ve_3}{x^2} D_{\o-1}^{32} + \CalO(x^{-3}) \right], \quad \o \in [\beta_-],
\end{align}
\end{subequations}
where
\begin{subequations}
\begin{align}
    & k_\o^{12,\pm} = |\EK_\o^{12,\pm}|, \ k_\o^{12} = k_\o^{12,+} - k_\o^{12,-}, \ \nu_\o = k_\o - k_{\o+1}, \\
    & \sigma_\o ^{12,\pm} = \sum_{\Box\in \EK_{\o}^{12,\pm} } c_\Box^{12,\pm}, \ \sigma_\o^{12} = \sigma_\o^{12,+} - \sigma_\o^{12,-}, \ D_\o^{12} = \frac{\ve_1}{2} k_\o  + \sigma_{\o-1} - \sigma_\o;   \\
    & k_\o^{32,\pm} = |\EK_\o^{32,\pm}|, \ k_\o^{32} = k_\o^{32,+} - k_\o^{32,-}, \ \nu_\o = k_\o - k_{\o+1}, \\
    & \sigma_\o ^{32,\pm} = \sum_{\Box\in \EK_{\o}^{32,\pm} } c_\Box^{32,\pm}, \ \sigma_\o^{32} = \sigma_\o^{32,+} - \sigma_\o^{32,-}, \ D_\o^{32} = \frac{\ve_3}{2} k_\o  + \sigma_{\o-1} - \sigma_\o. 
\end{align}
\end{subequations}

In order to have non-vanishing interaction between the particles in the different sector, one may deliberately modify the metric on the graded vector spaces $\BC^{N_+|N_-}$ \eqref{def:inner-product} to
\begin{align}\label{eq:metric-deform-12}
    \begin{pmatrix}
        + {\bf 1}_{N_+} & 0 \\ 0 & -\theta_1 {\bf 1}_{N_-}
    \end{pmatrix}.
\end{align}
The metric deformation breaks $U(N_+|N_-) \to U(N_+) \times U(N_-)$, which leads to the deformation of the commutation relation between the canonical momenta on the $\BC_{12}^2$:
\begin{align}
\begin{split}
    & [\rx_\alpha^+,p^+_{\rx,\alpha'}] = -\ve_1 \d_{\alpha\alpha'}; \quad \alpha,\alpha' = 1,\dots,N_+, \\
    & [\rx_\beta^-,p^-_{\rx,\beta'}] = \theta_1 \ve_1 \d_{\beta\beta'}; \quad \beta,\beta' = 1,\dots,N_-, \\
    & [\rx_\alpha^+,p^-_{\rx,\beta}] = 0 = [\rx_\beta^-,p^+_{\rx,\alpha}]; \quad \alpha=1,\dots,N_+, \ \beta = 1,\dots,N_-.
\end{split}
\end{align}
This is equivalent to modify the $\Omega$-deformation in the negative sector. We also deliberately modify the metric on the graded vector space $\BC^{M_+|M_-}$ to 
\begin{align}\label{eq:metric-deform-32}
    \begin{pmatrix}
        + {\bf 1}_{M_+} & 0 \\ 0 & - \theta_3 {\bf 1}_{M_-}
    \end{pmatrix}
\end{align}
The metric deformation breaks $U(M_+|M_-) \to U(M_+) \times U(M_-)$, which leads to the deformation of the commutation relation between the canonical momenta on the $\BC_{23}^2$ space: 
\begin{align}
\begin{split}
    & [\ry^+_\alpha,p^+_{\ry,\alpha'}] = -\ve_3 \d_{\alpha\alpha'}; \quad \alpha, \alpha' = 1,\dots,M_+, \\
    & [\ry^-_\beta,p^-_{\ry,\beta'}] = \theta_3 \ve_3 \d_{\beta\beta'}; \quad \beta,\beta' = 1,\dots,M_-, \\
    & [\ry_\alpha^+,p^-_{\ry,\beta}] = 0 = [\ry^-_\beta,p^+_{\ry,\alpha}]; \quad \alpha = 1,\dots,M^+, \ \beta = 1,\dots,M_-.
\end{split}
\end{align}
This is equivalent to modify the $\Omega$-deformation in the negative sector. 

The large $x$ expansion of the $qq$-characters can be be obtained through some straightforward though tedious computation. For later convenience, we will denote
\begin{align}
\begin{split}
    & P^{12,+}_\o=\ve_1\nu_{\omega}^{12}-a^+_{c^{-1}(\omega+1)},\ P^{12,-}_{\o}=\theta_1 \ve_1\nu_{\omega}^{12}+a^-_{c^{-1}(\omega+1)},\\ 
    & P^{32,+}_\o=\ve_3\nu_{\omega}^{32}-b^+_{c^{-1}(\omega+1)},\ P^{32,-}_\o=\theta_3\ve_3\nu_{\omega}^{32}+b^-_{c^{-1}(\omega+1)}. 
\end{split}
\end{align}
When $\o \in [\alpha_+]$:
\begin{align}
\begin{split}
    \frac{\CalX_{1,\o}(x)}{\BB_\o^{12}} = 
    &x+P_{\o}^{12,+}+\ve_1+\frac{1}{x}\left\{ \frac{ (P^{12,+}_\o)^2 - (a^+_{c^{-1}(\o+1)})^2 }{2} +\ve_1 D_\omega^{12} \right. \\
    & + \left[ \ve_3 \sum_{\omega'\in[\alpha_+]}\left[\ve_4 \nabla^\fq_\omega - P^{12,+}_{\o'}
    \nabla^\fz_{\omega'}\right]
    + \frac{\ve_3}{\theta_1}\sum_{\omega'\in[\alpha_-]}\left[ \theta_1 \ve_4 \nabla^\fq_\omega-P_{\o'}^{12,-} \nabla^\fz_{\omega'}\right] \right. \\
    & \left. \left. \qquad  + \ve_1 \sum_{\omega'\in[\beta_+]}\left[\ve_4 \nabla^\fq_\omega - P_{\o'}^{32,+} \nabla^\fz_{\omega'}\right]
    + \frac{\ve_1}{\theta_3}\sum_{\omega'\in[\beta_-]}\left[ \theta_3 \ve_4 \nabla^\fq_\omega-P_{\o'}^{32,-}\nabla^\fz_{\omega'}\right] \right] \log \BB_\o^{12} \right\} \\
    & + \CalO(x^{-2}).
\end{split}
\end{align}
When $\omega\in[\alpha_-]$: 
\begin{align}
\begin{split}
    \left[\frac{\CalX_{1,\omega}(x)}{\mathbb{B}_{\omega}^{12}}\right]^{-1} = 
    & x-P_\o^{12,-}+\theta_1 \ve_1+\frac{1}{x}\left\{ \frac{ (P^{12,+}_\o)^2 - (a^-_{c^{-1}(\o+1)})^2 }{2} -\theta \ve_1 D_\omega^{12} +\right. \\
    & - \left[ \ve_3 \sum_{\omega'\in[\alpha_+]}\left[\ve_4 \nabla^\fq_\omega - P^{12,+}_{\o'}
    \nabla^\fz_{\omega'}\right]
    + \frac{\ve_3}{\theta_1}\sum_{\omega'\in[\alpha_-]}\left[ \theta_1 \ve_4 \nabla^\fq_\omega-P_{\o'}^{12,-} \nabla^\fz_{\omega'}\right] \right. \\
    & \left. \left. \qquad  + \ve_1 \sum_{\omega'\in[\beta_+]}\left[\ve_4 \nabla^\fq_\omega - P_{\o'}^{32,+} \nabla^\fz_{\omega'}\right]
    + \frac{\ve_1}{\theta_3}\sum_{\omega'\in[\beta_-]}\left[ \theta_3 \ve_4 \nabla^\fq_\omega-P_{\o'}^{32,-}\nabla^\fz_{\omega'}\right] \right] \log \BB_\o^{12} \right\} \\
    & + \CalO(x^{-2}).
\end{split}
\end{align}
When $\o \in [\beta_+]$:
\begin{align}
\begin{split}
    \frac{\CalX_{3,\o}(x)}{\BB_\o^{32}} = 
    &x+P_{\o}^{32,+}+\ve_3+\frac{1}{x}\left\{ \frac{ (P^{32,+}_\o)^2 - (b^+_{c^{-1}(\o+1)})^2 }{2} +\ve_3 D_\omega^{32} \right. \\
    & + \left[ \ve_3 \sum_{\omega'\in[\alpha_+]}\left[\ve_4 \nabla^\fq_\omega - P^{12,+}_{\o'}
    \nabla^\fz_{\omega'}\right]
    + \frac{\ve_3}{\theta_1}\sum_{\omega'\in[\alpha_-]}\left[ \theta_1 \ve_4 \nabla^\fq_\omega-P_{\o'}^{12,-} \nabla^\fz_{\omega'}\right] \right. \\
    & \left. \left. \qquad  + \ve_1 \sum_{\omega'\in[\beta_+]}\left[\ve_4 \nabla^\fq_\omega - P_{\o'}^{32,+} \nabla^\fz_{\omega'}\right]
    + \frac{\ve_1}{\theta_3}\sum_{\omega'\in[\beta_-]}\left[ \theta_3 \ve_4 \nabla^\fq_\omega-P_{\o'}^{32,-}\nabla^\fz_{\omega'}\right] \right] \log \BB_\o^{32} \right\} \\
    & + \CalO(x^{-2}).
\end{split}
\end{align}
Finally when $\o \in [\beta_-]$:
\begin{align}
\begin{split}
    \left[\frac{\CalX_{3,\o}(x)}{\BB_\o^{32}} \right]^{-1} = 
    &x-P_{\o}^{32,-}+\theta_3\ve_3+\frac{1}{x}\left\{ \frac{ (P^{32,-}_\o)^2 - (b^-_{c^{-1}(\o+1)})^2 }{2} -\theta_3 \ve_3 D_\omega^{32} \right. \\
    & - \left[ \ve_3 \sum_{\omega'\in[\alpha_+]}\left[\ve_4 \nabla^\fq_\omega - P^{12,+}_{\o'}
    \nabla^\fz_{\omega'}\right]
    + \frac{\ve_3}{\theta_1}\sum_{\omega'\in[\alpha_-]}\left[ \theta_1 \ve_4 \nabla^\fq_\omega-P_{\o'}^{12,-} \nabla^\fz_{\omega'}\right] \right. \\
    & \left. \left. \qquad  + \ve_1 \sum_{\omega'\in[\beta_+]}\left[\ve_4 \nabla^\fq_\omega - P_{\o'}^{32,+} \nabla^\fz_{\omega'}\right]
    + \frac{\ve_1}{\theta_3}\sum_{\omega'\in[\beta_-]}\left[ \theta_3 \ve_4 \nabla^\fq_\omega-P_{\o'}^{32,-}\nabla^\fz_{\omega'}\right] \right] \log \BB_\o^{32} \right\} \\
    & + \CalO(x^{-2}).
\end{split}
\end{align}

We consider the following combination for the overall normalization constant,
\begin{align}
    \prod_{\o \in [\alpha_+]} \left[ \frac{\CalX_{1,\o}(x) }{\ve_1\BB_\o^{12}} \right] \prod_{\o \in [\alpha_-]} \left[ \theta_1 \frac{\CalX_{1,\o}(x) }{\ve_1\BB_\o^{12}} \right] \prod_{\o \in [\beta_+]} \left[ \frac{\CalX_{3,\o}(x) }{\ve_3\BB_\o^{32}} \right] \prod_{\o \in [\beta_-]} \left[ \theta_3 \frac{\CalX_{3,\o}(x) }{\ve_3\BB_\o^{32}} \right] .
\end{align}
Then, we obtain the following commuting Hamiltonians.

\emph{The first commuting Hamiltonian}:
\begin{align}
\begin{split}
    H_1 & = \frac{1}{\ve_1} \sum_{\o \in [\alpha_+]} c_{1,\o}^{12} + \frac{1}{\theta_1\ve_1} \sum_{\o \in [\alpha_-]} c_{1,\o}^{12} + \frac{1}{\ve_3} \sum_{\o \in [\beta_+]} c_{1,\o}^{32} + \frac{1}{\theta_3\ve_3} \sum_{\o \in [\beta_-]} c_{1,\o}^{32} \\
    & = \frac{1}{\ve_1} \sum_{\o \in [\alpha_+]} P_\o^{12,+} + \frac{1}{\theta_1\ve_1} \sum_{\o \in [\alpha_-]} P_\o^{12,-} + \frac{1}{\ve_3} \sum_{\o \in [\beta_+]} P_{\o}^{32,+} + \frac{1}{\theta_3\ve_3} \sum_{\o \in [\beta_-]} P_{\o}^{32,-}
\end{split}
\end{align}

\emph{The second commuting Hamiltonian}: 
\begin{align}\label{eq:H-2-raw}
\begin{split}
    H_2 = & \frac{1}{\ve_1} \sum_{\o \in [\alpha_+]} c_{2,\o}^{12} + \frac{1}{\theta_1\ve_1} \sum_{\o \in [\alpha_-]} c_{2,\o}^{12} + \frac{1}{\ve_3} \sum_{\o \in [\beta_+]} c_{2,\o}^{32} + \frac{1}{\theta_3\ve_3} \sum_{\o \in [\beta_-]} c_{2,\o}^{32} \\
    = & \frac{1}{2\ve_1} \sum_{\o \in [\alpha_+]} (P_\o^{12,+})^2 - \frac{1}{2\theta_1\ve_1} \sum_{\o \in [\alpha_-]} (P_\o^{12,-})^2 + \frac{1}{2\ve_3} \sum_{\o \in [\beta_+]} (P_{\o}^{32,+})^2 - \frac{1}{2\theta_3\ve_3} \sum_{\o \in [\beta_-]} (P_{\o}^{32,-})^2 \\
    & + \ve_3 \sum_{\omega\in[\alpha_+]}\left[\ve_4 \nabla^\fq_\omega - P^{12,+}_{\o}
    \nabla^\fz_{\omega}\right] \left[ \frac{1}{\ve_1} \log \BB^{12}_+ - \frac{1}{\theta_1\ve_1} \log \BB^{12}_- + \frac{1}{\ve_3} \log \BB^{32}_+ - \frac{1}{\theta_3\ve_3} \log \BB^{32}_- \right] \\
    & + \frac{\ve_3}{\theta_1}\sum_{\omega\in[\alpha_-]}\left[ \theta_1 \ve_4 \nabla^\fq_\omega-P_{\o}^{12,-} \nabla^\fz_{\omega}\right] \left[ \frac{1}{\ve_1} \log \BB^{12}_+ - \frac{1}{\theta_1\ve_1} \log \BB^{12}_- + \frac{1}{\ve_3} \log \BB^{32}_+ - \frac{1}{\theta_3\ve_3} \log \BB^{32}_- \right] \\
    & + \ve_1 \sum_{\omega\in[\beta_+]}\left[\ve_4 \nabla^\fq_\omega - P_{\o}^{32,+} \nabla^\fz_{\omega}\right] \left[ \frac{1}{\ve_1} \log \BB^{12}_+ - \frac{1}{\theta_1\ve_1} \log \BB^{12}_- + \frac{1}{\ve_3} \log \BB^{32}_+ - \frac{1}{\theta_3\ve_3} \log \BB^{32}_- \right] \\
    & + \frac{\ve_1}{\theta_3}\sum_{\omega\in[\beta_-]}\left[ \theta_3 \ve_4 \nabla^\fq_\omega-P_{\o}^{32,-}\nabla^\fz_{\omega}\right] \left[ \frac{1}{\ve_1} \log \BB^{12}_+ - \frac{1}{\theta_1\ve_1} \log \BB^{12}_- + \frac{1}{\ve_3} \log \BB^{32}_+ - \frac{1}{\theta_3\ve_3} \log \BB^{32}_- \right]
\end{split}
\end{align}
We will consider the following canonical transformation between the coordinate $\tq$ and its conjugate momentum $\tp = -\hbar\p_\tq$ which satisfies the commutation relation $[\tq,\tp] = \hbar$.
For a differentiable function $f(\fq)$, we have
\begin{align}
\begin{split}
    & \frac{1}{2}\frac{\p^2}{\p \tq^2} - f(\tq) \p_\tq \\
    & = \frac{1}{2} \left( \frac{\p}{\p\tq} - f(\tq) \right)^2 - \frac{1}{2} f(\tq)^2 - \frac{1}{2} f(\tq) \p_\tq + \frac{1}{2} \p_\tq f(\tq) \\
    & = \frac{1}{2} \left( \frac{\p}{\p\tq} - f(\tq) \right)^2 - \frac{1}{2} f(\tq)^2 + \frac{f'(\tq)}{2} .
\end{split}
\end{align}
The canonical transform gives the potential term in the Hamiltonian,
\begin{align}
\begin{split}
    \frac{1}{2} \sum_{\o\in[\alpha_+]} 
    & (\nabla^\fz_\o)^2 \ve_1 \ve_3 \left( \frac{1}{\ve_1} \log \BB^{12}_+ - \frac{1}{\theta_1\ve_1} \log \BB^{12}_- + \frac{1}{\ve_3} \log \BB^{32}_+ - \frac{1}{\theta_3\ve_3} \log \BB^{32}_- \right) \\ 
    & - \frac{\ve_3^2\ve_1^2}{\ve_1} \left[ \nabla^\fz_\o \left( \frac{1}{\ve_1} \log \BB^{12}_+ - \frac{1}{\theta_1\ve_1} \log \BB^{12}_- + \frac{1}{\ve_3} \log \BB^{32}_+ - \frac{1}{\theta_3\ve_3} \log \BB^{32}_- \right) \right]^2 \\
    +\frac{1}{2} \sum_{\o\in[\alpha_-]} 
    & \ve_1\ve_3 (\nabla^\fz_\o)^2 \left( \frac{1}{\ve_1} \log \BB^{12}_+ - \frac{1}{\theta_1\ve_1} \log \BB^{12}_- + \frac{1}{\ve_3} \log \BB^{32}_+ - \frac{1}{\theta_3\ve_3} \log \BB^{32}_- \right) \\
    & + \frac{\ve_3^2\ve_1^2}{\theta_1\ve_1} \left[ \nabla^\fz_\o \left( \frac{1}{\ve_1} \log \BB^{12}_+ - \frac{1}{\theta_1\ve_1} \log \BB^{12}_- + \frac{1}{\ve_3} \log \BB^{32}_+ - \frac{1}{\theta_3\ve_3} \log \BB^{32}_- \right) \right]^2 \\
    +\frac{1}{2} \sum_{\o\in[\beta_+]} 
    & \ve_1\ve_3 (\nabla^\fz_\o)^2 \left( \frac{1}{\ve_1} \log \BB^{12}_+ - \frac{1}{\theta_1\ve_1} \log \BB^{12}_- + \frac{1}{\ve_3} \log \BB^{32}_+ - \frac{1}{\theta_3\ve_3} \log \BB^{32}_- \right) \\
    & - \frac{\ve_3^2\ve_1^2}{\ve_3} \left[ \nabla^\fz_\o \left( \frac{1}{\ve_1} \log \BB^{12}_+ - \frac{1}{\theta_1\ve_1} \log \BB^{12}_- + \frac{1}{\ve_3} \log \BB^{32}_+ - \frac{1}{\theta_3\ve_3} \log \BB^{32}_- \right) \right]^2 \\
    +\frac{1}{2} \sum_{\o\in[\beta_-]} 
    & \ve_1\ve_3 (\nabla^\fz_\o)^2 \left( \frac{1}{\ve_1} \log \BB^{12}_+ - \frac{1}{\theta_1\ve_1} \log \BB^{12}_- + \frac{1}{\ve_3} \log \BB^{32}_+ - \frac{1}{\theta_3\ve_3} \log \BB^{32}_- \right) \\
    & + \frac{\ve_3^2\ve_1^2}{\theta_3\ve_3} \left[ \nabla^\fz_\o \left( \frac{1}{\ve_1} \log \BB^{12}_+ - \frac{1}{\theta_1\ve_1} \log \BB^{12}_- + \frac{1}{\ve_3} \log \BB^{32}_+ - \frac{1}{\theta_3\ve_3} \log \BB^{32}_- \right) \right]^2 \\
\end{split}
\end{align}
The canonical transformation combining with $\nabla^\kq_\o$ term gives 
the second Hamiltonian:
\begin{align}\label{eq:H_2}
\begin{split}
    H_2(\theta_1,\theta_3) = 
    & - \sum_{\alpha=1}^{N_+} \frac{1}{2}\frac{\p^2}{\p(\rx_\alpha^+)^2} - \sum_{\beta=1}^{M_+} \frac{g}{2} \frac{\p^2}{\p(\ry_\beta^+)^2} + \sum_{\alpha'=1}^{N_-} \frac{\theta_1}{2} \frac{\p^2}{\p(\rx_{\alpha'}^-)^2} + \sum_{\beta'=1}^{M_-} \frac{\theta_3g}{2}  \frac{\p^2}{\p(\ry_{\beta'}^-)^2} \\
    & + g(g+1) \sum_{1\leq \alpha<\alpha'\leq N_+} \wp (\rx_\alpha^+ - \rx_{\alpha'}^+) - \left( \frac{g^2}{\theta_1^3} + \frac{g}{\theta_1} \right) \sum_{1\leq \alpha<\alpha'\leq N_-} \wp (\rx_{\alpha}^- - \rx_{\alpha'}^-) \\
    & + \left( 1 + \frac{1}{g} \right) \sum_{1\leq \beta<\beta'\leq M_+} \wp(\ry_\beta^+ - \ry_{\beta'}^+) - \left( \frac{1}{g\theta_3^3} + \frac{1}{\theta_3} \right) \sum_{1\leq\beta<\beta'\leq M_-} \wp (\ry_\beta^- - \ry_{\beta'}^-) \\
    & + (g+1) \sum_{\alpha=1}^{N_+} \sum_{\beta=1}^{M_+} \wp (\rx_\alpha^+ - \ry_{\beta}^+)
    - \frac{1}{2} \left( \frac{g}{\theta_1} + \frac{g}{\theta_1^2\theta_3} + \frac{1}{\theta_3} + \frac{1}{\theta_3^2\theta_1} \right) \sum_{\alpha=1}^{N_-} \sum_{\beta=1}^{M_-} \wp (\rx_\alpha^- - \ry_{\beta}^-) \\
    & + \frac{1}{2} \left( \frac{g}{\theta_1} + \frac{g^2}{\theta_1} - g - \frac{g^2}{\theta_1^2} \right)   \sum_{\alpha=1}^{N_+} \sum_{\alpha'=1}^{N_-} \wp (\rx_\alpha^+ - \rx_{\alpha'}^-) \\
    & + \frac{1}{2} \left( 1 + \frac{1}{g\theta_3^2} - \frac{1}{\theta_3} - \frac{1}{g\theta_3} \right) \sum_{\beta=1}^{M_+} \sum_{\beta'=1}^{M_-} \wp (\ry_\beta^+ - \ry_{\beta'}^-) \\
    & + \frac{1}{2} \left( g + \frac{1}{\theta_3^2} - \frac{1}{\theta_3} - \frac{g}{\theta_3} \right) \sum_{\alpha=1}^{N_+} \sum_{\beta=1}^{M_-} \wp (\rx_\alpha^+ - \ry_{\beta}^-) \\
    & + \frac{1}{2} \left( 1- \frac{1}{\theta_1} + \frac{g}{\theta_1^2} - \frac{g}{\theta_1} \right) \sum_{\alpha=1}^{N_-} \sum_{\beta=1}^{M_+} \wp (\ry_{\beta}^+-\rx_\alpha^-) 
\end{split}
\end{align}
where $g = \frac{\ve_3}{\ve_1}$. The elliptic modulus of the Weierstrass $\wp$-function is identified with the complexified gauge coupling
\begin{equation}
    \tau = \frac{4\pi\ri}{g_{\rm YM}^2} + \frac{\vartheta}{2\pi} = 2\ell + 2\ri \delta.     \label{eq:cmplx_coupling}
\end{equation}

There are some special cases of the deformation parameter that we would like to address here: 

\begin{enumerate}
    \item In the undeformed case $\theta_1=\theta_3=1$, the Hamiltonian $H_2$ \eqref{eq:H_2} describes two set of edCM systems
    \begin{align}
        H_2(\theta_1=\theta_3=1) = {\rm H}_{N_+,M_+}(\rx^+,\ry^+;-g) - {\rm H}_{N_-,M_-}(\rx^-,\ry^-;-g)
    \end{align}
    without any interaction in between. 
    \item When $\theta_1=\theta_3=-1$, the system becomes a single edCM system defined on two ordinary groups
    \begin{align}
        H_2(\theta_1=\theta_3=-1) = {\rm H}_{N_++N_-,M_++M_-}(\rx^\pm,\ry^\pm;-g).
    \end{align}
    This agrees with the metric deformations \eqref{eq:metric-deform-12} and \eqref{eq:metric-deform-32}. 
    \item In the case $\theta_1 = -g = \theta_3^{-1}$. The eqCM system proposed in \cite{Berntson:2023prh} is recovered,
    \begin{align}
        H_2(\theta_1=-g=\theta_3^{-1}) = {\rm H}_{N_+,M_+,M_-,N_-}(\rx^+,\ry^+,\ry^-+\ri\d,\rx^-+\ri\d;-g) .
    \end{align}
    One further shifts $\rx^- \to \rx^- -\ri \d$ and $\ry^- \to \ry^- - \ri \d$ to obtain the eqCM system of particles with different chirality. 
    \item By setting $\theta_1=g=\theta_3^{-1}$, we recover the two copies of non-interacting edCM similar to the previous case $\theta_1=\theta_3=1$:
    \begin{align}
        H_2(\theta_1=g=\theta_3^{-1}) = {\rm H}_{N_+,M_+}(\rx^+,\ry^+;-g) - {\rm H}_{M_-,N_-}(\ry^-,\rx^-;-g).
    \end{align}
\end{enumerate}

\subsection{Trigonometric limit}
The trigonometric limit of both eCM and edCM are obtained in the $\d \to \infty$ limit
\[
    \lim_{\d \to \infty} \wp(z) = \frac{1}{\sin^2z}, 
\]
which we denote as trigonometric Calogero-Moser (tCM) and trigonometric double Calogero-Moser (tdCM) respetively. 
In the context of gauge theory, this corresponds to the weak coupling limit~\eqref{eq:cmplx_coupling}.

The eqCM system with all particles sharing the same chirality is no different of an edCM system. The trigonometric limit of eqCM Hamiltonian $H_2(\theta_1 =\theta_3^{-1} = g)$ \eqref{eq:H_2} of a single chirality is the tdCM of $L=N_++N_-+M_++M_-$ particles. 
In the case of eqCM system consisting particles of different chirality, the potential term between particles with different chirality vanishes in the trigonometric limit:
\[
    \lim_{\d\to \infty} \wp(z+\ri\d) = 0.
\] 
The operator $H_2(\theta_1 =\theta_3^{-1} = g)$ is reduced to a sum of two commuting tdCM Hamiltonians. 
Instead, one can consider the trigonometric limit of the Hamiltonian $H_2(\theta_1,\theta_3)$ in \eqref{eq:H_2} with general deformation parameter $\theta_1,\theta_3$. 

\section{Summary and Discussion} \label{sec:discussion}

We have proven the integrability of eqCM system by explicitly constructing the corresponding Dunkl operator. 
We have then established the Bethe/Gauge correspondence of the eqCM system with supergroup folded instanton. 
The supergroups must be deliberately broken to have non-vanishing interaction between particles with opposite chirality. By introducing the full-type co-dimensional two defect, we reconstruct the quantum Hamiltonian of the eqCM system from supergroup folded instanton through its $qq$-character.

Let us end with discussing a few loose ends in this note and commenting on a few future directions.

\begin{itemize}
    \item A trigonometric triple Calogero-Moser (ttCM) system is proposed in \cite{Gaiotto:2020dsq}. 
    The Hamiltonian \eqref{eq:H_2} does not (seem to) recover the elliptic version of the triple Calogero-Moser system by proper choosing of the deformation parameters $(\theta_1,\theta_3)$. 
    \item Calogero-Moser systems can be defined on any root system of Lie (super)algebra. 
    From this point of view, it is natural to ask what is the root system corresponding to the eqCM system defined in \cite{Berntson:2023prh}.
    A plausible candidate is that associated with $\BZ_2\times \BZ_2$-graded superalgebra \cite{Stoilova:2023nam,Stoilova:2024wlf}. 
    \item Eigenfunctions of the tCM system are given by Jack polynomials \cite{Kimura:2022zsx}. The surface defect partition function, an infinite series in fractional coupling $(\kq_\o)_{\o=0,\dots,L-1}$, is reduced to Jack polynomial through a proper Higgsing condition on the Coulomb moduli. 
    The eigenfunction of the tqCM system, which is the defect folded instanton partition function, would be a double-graded generalization of Jack polynomial under proper Higgsing condition.  
    \item The gauge origami we discussed in this paper is a generalization of the 4d $\CalN=2$ supersymmetric gauge theory. The D(-1)-D3-$\overline{\rm D3}$ brane setup Table.~\ref{tab:GO} can be T-dualized to D0-D4-$\overline{\rm D4}$ brane system, whose worldvolume becomes generalization of 5d $\CalN=1$ supersymmetric gauge theory. 
    The corresponding integrable system is relativistic (trigonometric for momentum variables) \cite{Lee:2023wbf}. 
\end{itemize}

\appendix

\section{Superalgebra and supermatrix}\label{Superalgebra}
In this short appendix we also specify our conventions of superalgebra and supermatrices used in the main text.
See also \cite{Kac:1977em,Quella:2013oda} for details.
\subsection{Superalgebra}

A superalgebra is a $\bbZ_2$-graded algebra. It is an algebra over a commutative ring or field $\mathbb{K}$ with decomposition into even and odd elements and a multiplication operator that respects the grading. A superalgebra over $\mathbb{K}$ is defined by direct sum decomposition
\begin{align}
    A=A_0\oplus A_1
\end{align}
with a bilinear multiplication $A\times A\to A$ such that 
\begin{align}
    A_i\times A_j\subseteq A_{i+j}.
\end{align}
A parity is assigned to every element $x\in A$, denoted as $|x|$, is either $0$ or $1$ depending on whether $x$ is in $A_0$ or $A_1$. We define supercommutator by
\begin{align}\label{super odd}
    [x,y]=xy-(-1)^{|x||y|}yx.
\end{align}
A superalgebra $A$ is said to be supercommutative if 
\begin{align}
    [x,y]=0\quad \forall x,y\in A.
\end{align} 

\subsection{Supermatrix}

A supermatrix is a $\bbZ_2$-graded analog of ordinary matrix. It is a $2\times2$ block matrix with entries in superalgebra $R$. $R$ can be either commutative superalgebra (e.g. Grassmannian algebra) or ordinary field. A supermatrix of dimension $(r|s)\times(p|q)$ is a matrix
\begin{align}\label{supermatrix blockform}
    X=\begin{pmatrix}
    X_{00} & X_{01} \\
    X_{10} & X_{11}
    \end{pmatrix}
\end{align}
with $r+s$ rows and $p+q$ columns. The block matrices $X_{00}$ and $X_{11}$ consist solely of even graded element in $R$, while
$X_{01}$ and $X_{10}$ consist solely of odd graded elements in $R$. 
If $X$ is a square matrix, its supertrace is defined as
\begin{align}
    \operatorname{str}(X)=\operatorname{tr}(X_{00})-\operatorname{tr}(X_{11}).
\end{align}
For an invertible supermatrix over commutative superalgebra, its superdeterminant (Berezinian) is defined as
\begin{align}
    \operatorname{sdet} X =\det(X_{00}-X_{01}X^{-1}_{11}X_{10})\det(X_{11})^{-1} = \det (X_{00}) \det(X_{11} - X_{10}X_{00}^{-1}X_{01})^{-1}.
\end{align}

\subsection{Gauge origami supermatrix}

Here we write down explicitly the supergroup gauge origami maps $B_a$, $I_A$, and $J_A$ , $a\in \underline{4}$, $A\in \underline{6}$, in the supermatrix form: 
\begin{subequations}
\begin{align}
    B_a & = 
    \begin{pmatrix}
        B_{a,00} & B_{a,01} \\ B_{a,10} & B_{a,11}
    \end{pmatrix} \\
    I_A & = \begin{pmatrix}
        I_{A,00} & I_{A,01} \\ I_{A,10} & I_{A, 11}
    \end{pmatrix} \\
    J_A & = \begin{pmatrix}
        J_{A,00} & J_{A,01} \\ J_{A,10} & J_{A, 11}
    \end{pmatrix} 
\end{align}
\end{subequations}
The components of the off-diagonal block matrix are Grassmannian numbers. 

The real moment map of the gauge origami \eqref{mu=0} can be expressed in the terms of the supermatrices,
\begin{align}\label{eq:real-moment-supermatrix}
\begin{split}
    & \mu_\BR = \\
    & \sum_{a\in \underline{4}}
    \begin{pmatrix}
    [B_{a,00},B_{a,00}^\dagger] + [B_{a,01},B_{a,01}^\dagger] & B_{a,00}B_{a,10}^\dagger + B_{a,01}B_{a,11}^\dagger - B_{a,10}B_{1,00}^\dagger - B_{a,11}B_{a,01}^\dagger \\
    B_{a,10}B_{1,00}^\dagger + B_{a,11}B_{a,01}^\dagger - B_{a,00}B_{1,10}^\dagger - B_{a,01}B_{a,11}^\dagger & [B_{a,10},B_{a,10}^\dagger] + [B_{a,11},B_{a,11}^\dagger]
    \end{pmatrix} \\
    & + \sum_{A\in \underline{6}}
    \begin{pmatrix}
    I_{A,00}I_{A,00}^\dagger + I_{A,01}I_{A,01}^\dagger & I_{A,00}I_{A,10}^\dagger + I_{A,01}I_{A,11}^\dagger \\
    I_{A,10}I_{A,00}^\dagger + I_{A,11}I_{A,01}^\dagger & I_{A,10}I_{A,01}^\dagger + I_{A,11}I_{A,11}^\dagger
    \end{pmatrix} - 
    \begin{pmatrix}
    J_{A,00}^\dagger J_{A,00} + J_{A,10}^\dagger J_{A,10} & J_{A,00}^\dagger J_{A,01} + J_{A,10}^\dagger J_{A,11} \\ 
    J_{A,01}^\dagger J_{A,00} + J_{A,11}^\dagger J_{A,10} & J_{A,01}^\dagger J_{A,01} + J_{A,11}^\dagger J_{A,11}
    \end{pmatrix} 
\end{split}
\end{align}

\section{List of relevant mathematical functions}

In this appendix, we provide some relevant details about the mathematical functions used in the main text.

\subsection{Random partition}

A partition is defined as a way of expressing a non-negative integer $n$ as a summation over other non-negative integers. Each partition can be labeled by a Young diagram $\lambda=(\lambda_1 \ge \lambda_2 \ge \dots \ge \lambda_{\ell(\lambda)} > 0)$ with $\lambda_i\in\mathbb{N}$ such that 
\begin{equation}
n=|\lambda|=\sum_{i=1}^{\ell(\lambda)}\lambda_i,
\end{equation} 
where $\ell(\lambda)$ denotes the number of rows in $\lambda$, i.e. $\ell(\lambda) = \lambda_1^\text{T}$
We define the generating function of such a partition as
\begin{subequations}\label{phi}
\begin{align}
&\sum_{\lambda}\mathfrak{q}^{|\lambda|}=\frac{1}{(\mathfrak{q};\mathfrak{q})_\infty},\quad(\mathfrak{q};\mathfrak{q})_\infty=\prod_{n=1}^\infty\left(1-\mathfrak{q}^n\right); \\
&\sum_{\lambda}t^{\ell(\lambda)}\mathfrak{q}^{|\lambda|}=\frac{1}{(\mathfrak{q}t;\mathfrak{q})_\infty};\quad (\mathfrak{q}t;\mathfrak{q})_\infty=\prod_{n=1}^\infty(1-t\mathfrak{q}^n),
\end{align}
\end{subequations}
for $|\fq| < 1$.
Here we also define the $\mathfrak{q}$-shifted factorial (the $\mathfrak{q}$-Pochhammer symbol) as
\begin{equation}\label{q-Pochhammer}
(z;\fq)_n=\prod_{m=0}^{n-1}(1-z\fq^m).
\end{equation}

\subsection{The elliptic functions}

Here we fix our notation for the elliptic functions. The so-called Dedekind eta function is denoted as
\begin{equation}\label{eta}
\eta(\tau)=e^{\frac{\pi i\tau}{12}}(\mathfrak{q};\mathfrak{q})_\infty,
\end{equation}
where $\mathfrak{q}=\exp \left( 2 \pi i \tau \right)$.
The first Jacobi theta function is denoted as:
\begin{equation}\label{theta}
\theta_{11}(z;\tau)=ie^{\frac{\pi i\tau}{4}}z^{\frac{1}{2}}(\mathfrak{q};\mathfrak{q})_\infty (\mathfrak{q}z;\mathfrak{q})_\infty(z^{-1};\mathfrak{q})_\infty,
\end{equation} 
whose series expansion 
\begin{equation}\label{theta2}
\theta_{11}(z;\tau)=i\sum_{r\in\mathbb{Z}+\frac{1}{2}}(-1)^{r-\frac{1}{2}}z^re^{\pi i\tau r^2}=i\sum_{r\in\mathbb{Z}+\frac{1}{2}}(-1)^{r-\frac{1}{2}}e^{rx}e^{\pi i\tau r^2},
\end{equation}
implies that it obeys the heat equation
\begin{equation}
\frac{1}{\pi i}\frac{\partial}{\partial\tau}\theta_{11}(z;\tau)=(z\partial_z)^2\theta_{11}(z;\tau).
\end{equation}
The Weierstrass $\wp$-function
\begin{equation}\label{Def:p-function}
\wp(z) = \frac{1}{z^2}+\sum_{p, q \ge 0} \left\{\frac{1}{(z+p+q\tau)^2}-\frac{1}{(p+q\tau)^2}\right\},
\end{equation}
is related to the theta and eta functions by
\begin{equation}
\wp(z;\tau)=-(z\partial_z)^2\log\theta_{11}(z;\tau)+\frac{1}{\pi i}\partial_\tau\log\eta(\tau).
\end{equation}

\subsection{Higher rank theta function}
Let us also define
\begin{equation}
\Theta_{\widehat{A}_{N-1}}(\vec{z};\tau)=\eta(\tau)^{N}\prod_{\alpha>\beta}\frac{\theta_{11}(z_\alpha/z_\beta;\tau)}{\eta(\tau)}
\end{equation}
as the rank $N-1$ theta function, which also satisfies the generalized heat equation
\begin{equation}\label{heat}
N\frac{\partial}{\partial\tau}\Theta_{\widehat{A}_{N-1}}(\vec{z};\tau)=\pi i\Delta_{\vec{z}}\Theta_{\widehat{A}_{N-1}}(\vec{z};\tau),
\end{equation}
with the $N$-variable Laplacian:
\begin{align}
    \Delta_{\vec{z}}=\sum_{\omega=0}^{N-1}(z_\omega\partial_{z_\omega})^2.
\end{align}

\subsection{The orbifolded partition}

For the purpose in the main text, we consider the orbifolded coupling 
\begin{align}
    \mathfrak{q}=\prod_{\omega=0}^{N-1}\mathfrak{q}_\omega;\quad \mathfrak{q}_{\omega+N}=\mathfrak{q}_\omega,
\end{align}
and
\begin{align}
    \mathfrak{q}_\omega=\frac{z_\omega}{z_{\omega-1}};\quad z_{\omega+N}=\mathfrak{q}z_{\omega}.
\end{align}
We also consider the orbifolded version of the generating function of partitions $(\mathfrak{q};\mathfrak{q})_\infty^{-1}$ in \eqref{phi}. Given a finite partition $\lambda=(\lambda_1,\dots,\lambda_{\ell(\lambda)})$, we define
\begin{equation}
\mathbb{Q}^{\lambda}_\omega
=\prod_{j=1}^{\lambda_1}\mathfrak{q}_{\omega+1-j}^{\lambda_j^\text{T}}
=\prod_{i=1}^{\ell(\lambda)}\frac{z_\omega}{z_{\omega-\lambda_{i}}}.
\end{equation}
where the transposed partition is denoted by $\lambda^\text{T}$.
The summation over all possible partitions is given by
\begin{equation}\label{Q-form}
\mathbb{Q}_\omega=\sum_{\lambda}\mathbb{Q}^{\lambda}_\omega=\sum_{\lambda}\prod_{i=1}^{\ell(\lambda)}\left(\frac{z_\omega}{z_{\omega-\lambda_{i}}}\right)=\sum_{l_0,\dots,l_{N-1},l\geq0}\prod_{\alpha=1}^{N-1}\left(\frac{z_\omega}{z_{\alpha}}\right)^{l_\alpha}\mathfrak{q}^l.
\end{equation}
The function $\mathbb{Q}(\vec{z};\tau)$ is the orbifolded version of the generating function of partitions \eqref{phi}, 
\begin{align}\label{Q}
\mathbb{Q}(\vec{z};\tau)
&=\prod_{\omega=0}^{N-1}\mathbb{Q}_\omega(\vec{z};\tau) \nonumber\\
&=\prod_{N-1 \geq \alpha>\beta \geq 0}\frac{1}{(\frac{z_\alpha}{z_\beta};\mathfrak{q})_\infty(\mathfrak{q}\frac{z_\beta}{z_\alpha};\mathfrak{q})_\infty}\prod_{\alpha=0}^{N-1}\frac{1}{(\mathfrak{q};\mathfrak{q})_\infty} \nonumber\\
&=\prod_{N-1\geq\alpha>\beta\geq0}\frac{\mathfrak{q}^{1/12}\eta(\tau)\sqrt{z_\alpha/z_\beta}}{\theta_{11}(z_\alpha/z_\beta;\tau)}\times\left[\frac{\mathfrak{q}^{1/24}}{\eta(\tau)}\right]^N \nonumber\\
&=\left[\eta(\tau)^{-N}\prod_{N-1\geq\alpha>\beta\geq0}\frac{\eta(\tau)}{\theta_{11}(z_\alpha/z_\beta;\tau)}\right]\frac{\mathfrak{q}^{N^2/24}}{\vec{z}^{\vec{\rho}}} \nonumber\\
&=\frac{1}{\Theta_{\widehat{A}_{N-1}}(\vec{z};\tau)}\frac{\mathfrak{q}^{N^2/24}}{\vec{z}^{\vec{\rho}}},
\end{align}
where $\vec{\rho}$ is the Weyl vector associated with $\mathfrak{sl}(N)$ Lie algebra, whose entries are given as
\begin{equation}
\vec{\rho}=(\rho_0,\dots,\rho_{N-1});\quad\rho_\omega=\omega-\frac{N-1}{2};\quad |\vec{\rho}|^2=\sum_{\omega=0}^{N-1}\rho_\omega^2=\frac{N(N^2-1)}{12};\quad\vec{z}^{\vec{\rho}}=\prod_{\omega=0}^{N-1}z_\omega^{\rho_\omega}.
\end{equation}
Using eq.~\eqref{heat}, it is easy to prove that the $\mathbb{Q}$-function satisfies
\begin{equation}\label{Heat eq for Q}
0=\sum_{\omega}\nabla^{\mathfrak{q}}_\omega\log\mathbb{Q}-\frac{1}{2}\Delta_{\vec{z}}\log\mathbb{Q}+\frac{1}{2}\sum_\omega(\nabla^z_\omega\log\mathbb{Q})^2,
\end{equation}
with 
\begin{equation}
\sum_\omega\nabla_\omega^\mathfrak{q}=N\nabla^\mathfrak{q}+\vec{\rho}\cdot{\nabla}^{\vec{z}} .
\end{equation}

\newpage
\bibliographystyle{utphys}
\bibliography{DDCM}

\end{document}